\documentclass[apj]{emulateapj}
\usepackage{apjfonts}

\newcommand{\muHz}{\mbox{$\mu$Hz}}

\received{August 20, 2008}
\accepted{November 4, 2008}

\slugcomment{Accepted for publication in the Astrophysical Journal}

\shorttitle{Whole Earth Telescope Observations of GD358}
\shortauthors{Provencal et al.}

\begin{document}


\title{2006 Whole Earth Telescope Observations of GD358: A New Look at
  the Prototype DBV}


\author{J.~L.~Provencal\altaffilmark{1,2},
M.~H.~Montgomery\altaffilmark{2,3},
A.~Kanaan\altaffilmark{4},
H.~L.~Shipman\altaffilmark{1},
D.~Childers\altaffilmark{5},
A.~Baran\altaffilmark{6},
S.~O.~Kepler\altaffilmark{7},
M.~Reed\altaffilmark{8},
A.~Zhou\altaffilmark{8},
J.~Eggen\altaffilmark{8},
T.~K.~Watson\altaffilmark{9},
D.~E.~Winget\altaffilmark{3},
S.~E.~Thompson\altaffilmark{1,2},
B.~Riaz\altaffilmark{1},
A.~Nitta\altaffilmark{10},
S.~J.~Kleinman\altaffilmark{10},
R.~Crowe\altaffilmark{11},
J.~Slivkoff\altaffilmark{11},
P.~Sherard\altaffilmark{11},
N.~Purves\altaffilmark{11},
P.~Binder\altaffilmark{11},
R.~Knight\altaffilmark{11},
S.~-L. Kim\altaffilmark{12},
Wen-Ping~Chen\altaffilmark{13},
M.~Yang\altaffilmark{13},
H.~C.~Lin\altaffilmark{13},
C.~C.~Lin\altaffilmark{13},
C.~W.~Chen\altaffilmark{13},
X.~J.~Jiang\altaffilmark{14},
A.~V.~Sergeev\altaffilmark{15},
D.~Mkrtichian\altaffilmark{16},
E.~Janiashvili\altaffilmark{15},
M.~Andreev\altaffilmark{15},
R.~Janulis\altaffilmark{17},
M.~Siwak\altaffilmark{18},
S.~Zola\altaffilmark{6,18},
D.~Koziel\altaffilmark{18},
G.~Stachowski\altaffilmark{6},
M.~Paparo\altaffilmark{19},
Zs.~Bognar\altaffilmark{19},
G.~Handler\altaffilmark{20},
D.~Lorenz\altaffilmark{20},
B.~Steininger\altaffilmark{20},
P.~Beck\altaffilmark{20},
T.~Nagel\altaffilmark{21},
D.~Kusterer\altaffilmark{21},
A.~Hoffman\altaffilmark{21},
E.~Reiff\altaffilmark{21},
R.~Kowalski\altaffilmark{21},
G.~Vauclair\altaffilmark{22},
S.~Charpinet\altaffilmark{22},
M.~Chevreton\altaffilmark{23},
J.~E.~Solheim\altaffilmark{24},
E.~Pakstiene\altaffilmark{25},
L.~Fraga\altaffilmark{4},
and
J.~Dalessio\altaffilmark{1,2}
}


\altaffiltext{1}{University of Delaware, Department of Physics and Astronomy
Newark, DE 19716; jlp@udel.edu}
\altaffiltext{2}{Delaware Asteroseismic Research Center, Mt. Cuba Observatory, Greenville, DE 19807}
\altaffiltext{3}{Department of Astronomy, University of Texas, Austin, TX 78712; mikemon@rocky.as.utexas.edu}
\altaffiltext{4}{Departamento de F\'{\i}sica Universidade Federal de Santa Catarina, C.P. 476, 88040-900, Florian{\'o}polis, SC, Brazil; ankanaan@gmail.com}
\altaffiltext{5}{Department of Math and Science, Delaware County Community College, 901 S. Media Rd, Media, PA 19063; dpc@udel.edu}
\altaffiltext{6}{Mount Suhora Observatory, Cracow Pedagogical University, Ul. Podchorazych 2, 30-084 Krakow, Poland; zola@astro1.as.ap.krakow.pl}
\altaffiltext{7}{Instituto de F\'{\i}isica UFRGS, C.P. 10501, 91501-970 Porto Alegre,RS, Brazil; kepler@if.ufrgs.br}
\altaffiltext{8}{Missouri State University and Baker Observatory, 901S. National, Springfield, MO 65897; MikeReed@missouristate.edu}
\altaffiltext{9}{Southwestern University, Georgetown, TX; tkw@southwestern.edu}
\altaffiltext{10}{Gemini Observatory, Northern Operations Center, 670 North A'ohoku Place, Hilo, HI 96720; atsuko.nittakleinman@gmail.com}
\altaffiltext{11}{University of Hawaii, Hilo, 96720; rcrowe@hubble.uhh.hawaii.edu}
\altaffiltext{12}{Korea Astronomy and Space Science Institute, Daejeon 305-348, Korea; slkim@kasi.re.kr}
\altaffiltext{13}{Lulin Observatory, National Central University, Taiwan; wchen@astro.ncu.edu.tw}
\altaffiltext{14}{National Astronomical Observatories, Academy of Sciences, Beijing 100012, PR China; xjjiang@bao.ac.cn}
\altaffiltext{15}{Ukrainian National Academy of Sciences, Main Astronomical Observatory, Golosiiv, Kiev 022 252650; sergeev@terskol.com }
\altaffiltext{16}{Odessa Astronomical Observatory, Odessa National University, T.G Shevchenko Park, Odessa 65014; david@arcsec.sejong.ac.kr}
\altaffiltext{17}{Institute of Theoretical Physics and Astronomy, Vilnius University,
Vilnius, Lithuania; jr@itpa.lt}
\altaffiltext{18}{Astronomical Observatory, Jagiellonian University, ul. Orla 171, 30-244 Cracow, Poland}
\altaffiltext{19}{Konkoly Observatory, P.O. Box 67, H-1525 Budapest XII, Hungary; paparo@konkoly.hu }
\altaffiltext{20}{Institut f\"ur Astronomie Universit\"at Wien, T\"urkenschanzstrasse 17, 1180, Austria; handler@astro.univie.ac.at}
\altaffiltext{21}{Institut f\"ur Astronomie und Astrophysik, Universi\"at T\"ubingen, Sand 1, 72076 T\"ubingen, Germany; nagel@astro.uni-tuebingen.de }
\altaffiltext{22}{Laboratoire d'Astrophysique de Toulouse-Tarbes, Universit\'e de Toulouse, CNRS, 14 avenue 
Edouard Belin, F314000 Toulouse, France; gerardv@srvdec.obs-mip.fr }
\altaffiltext{23}{ Observatoire de Paris, LESIA, 92195 Meudon, France}
\altaffiltext{24}{Institute of Theoretical Astrophysics, University of Oslo, PO Box 1029, Oslo, Norway; janerik@phys.uit.no}
\altaffiltext{25}{Institute of Theoretical Physics and Astronomy, Astronomical Observatory, Gostauto 12, Vilnius LT-2600, Lithuania; erika@itpa.lt}


\clearpage
\newpage

\begin{abstract}
  We report on the analysis of 436.1 hrs of nearly continuous
  high-speed photometry on the pulsating DB white dwarf GD358 acquired
  with the Whole Earth Telescope (WET) during the 2006 international
  observing run, designated XCOV25.  The Fourier transform (FT) of the
  light curve contains power between 1000 to 4000 \muHz, with the
  dominant peak at 1234 \muHz. We find 27 independent frequencies
  distributed in 10 modes, as well as numerous combination
  frequencies.  Our discussion focuses on a new asteroseismological
  analysis of GD358, incorporating the 2006 data set and drawing on 24
  years of archival observations.  Our results reveal that, while the
  general frequency locations of the identified modes are consistent
  throughout the years, the multiplet structure is complex and cannot
  be interpreted simply as $l$=1 modes in the limit of slow rotation.
  The high $k$ multiplets exhibit significant variability in
  structure, amplitude and frequency.  Any identification of the $m$
  components for the high $k$ multiplets is highly suspect.  The $k$=9
  and 8 modes typically do show triplet structure more consistent with
  theoretical expectations.  The frequencies and amplitudes exhibit
  some variability, but much less than the high $k$ modes. Analysis of
  the $k$=9 and 8 multiplet splittings from 1990 to 2008 reveal a
  long-term change in multiplet splittings coinciding with the 1996
  \emph{sforzando} event, where GD358 dramatically altered its
  pulsation characteristics on a timescale of hours.  We explore
  potential implications, including the possible connections between
  convection and/or magnetic fields and pulsations.  We suggest future
  investigations, including theoretical investigations of the
  relationship between magnetic fields, pulsation, growth rates, and
  convection.
\end{abstract}





\section{Introduction}

Asteroseismology of stellar remnants is traditionally the study of the
interior structure of pulsating white dwarfs and subdwarfs as revealed
by global stellar oscillations. The oscillations allow a view beneath
the photosphere, and contain information about basic physical
parameters, such as mass, rotation rate, internal transition profiles,
and compositional structure.  This information (see for example:
Nather et al. 1990, Winget et al.  1991, Winget et al. 1994, Kepler et
al. 2000, Kanaan et al. 2005) provides important constraints on fields
ranging from stellar formation and evolution, chemical evolution in
our galaxy, the age of the galactic disk, the physics of Type Ia
supernovae, and neutrino physics (Winget et al. 2004).

Asteroseismology is now expanding its focus to attack problems that at
first consideration may not seem best suited for the technique.
Convection remains one of the largest sources of theoretical
uncertainty in our understanding of stars.  This lack of understanding
leads to considerable systematic uncertainties in such important
quantities as the ages of massive stars where convective overshooting
is important (Di Mauro et al. 2003) and the temperatures and cooling
ages of white dwarfs (Prada et al. 2002, Wood 1992). One important
early result from the Canadian asteroseismology mission MOST is the
difficulty in detecting predicted oscillations in Procyon A, implying
possibly serious incompleteness in our understanding of stars even
slightly different from the Sun (Matthews et al. 2004, Marchenko
2008).

Montgomery (2005) shows how precise observations of variable star
light curves can be used to characterize the convection zone in a
pulsating star. Montgomery bases his approach on important analytical
(Goldreich \& Wu 1999, Wu 2001) and numerical precursor calculations
(Brickhill 1992). The method is based on three assumptions: 1) the
flux perturbations are sinusoidal below the convection zone, 2) the
convection zone is so thin that local angular variations of the
nonradial pulsations may be ignored: i.e., we treat the pulsations
locally as if they were radial, and 3) the convective turnover time is
short compared with the pulsation timescale, so the convection zone
can be assumed to respond instantaneously.  Using high signal-to-noise
light curves to model nonlinear effects, this approach can
observationally determine the convective time scale $\tau_0$, a
temperature dependence parameter N, and, together with an independent
$T_{\rm eff}$ determination, the classical convective efficiency
parameter (the mixing length ratio) $\alpha$.

Montgomery (2005) applies this theoretical construct to two large
amplitude, monoperiodic white dwarf variables where it is possible to
fold long light curves to obtain high signal to noise pulse shapes.
The test candidates are a hydrogen atmosphere DAV (G29-38) and a
helium atmosphere DBV (PG1351+489).  G29-38 is a well studied DAV
known for the complexity of its pulsations (Kleinman et al. 1998).
However, G29-38 was nearly monoperiodic during the 1988 Whole Earth
Telescope (WET) campaign (Winget et al. 1990).  PG1351+489 is
dominated by a single mode and its harmonics, and was WET target in
1995 (Alves et al. 2003). Using folded light curves, Montgomery (2005)
obtained results for these two stars which are consistent with mixing
length theory (MLT) and other calculations of convective transport.

This significant theoretical advance offers the first empirical
determination of convection parameters in stars other than the Sun.
However, a result from one star in each white dwarf instability strip
provides an insufficient basis for global statements about the nature
of the convection zones for all white dwarfs. The next logical step is
to map a population spanning a range of temperatures and masses across
each instability strip, enabling us to determine the depth of their
convection zones as a function of $T_{\rm eff}$ and log g.

PG1351+489 and G29-38 are examples of relatively rare, large
amplitude, monoperiodic pulsators where it is possible to fold long
light curves to obtain a high signal to noise pulse shapes.
Simulations by Montgomery (2006) show that this approach is not
sufficient for the more common pulsators demonstrating variable
complexity in their pulsation spectra.  In these cases, we require at
least 5 hours of very high signal to noise photometry (S/N
$\approx1000$) coupled with accurate real time frequency, amplitude,
and phase information for the pulsations present in the high S/N light
curve.  The frequency, amplitude, and phase information, provided by a
long timebase, multisite observing run, is used to fit the high S/N
light curve and extract the convection parameters.  The criteria for
such a candidate star include nonlinear pulsations, a fairly bright
target, and prior knowledge of the $l$ and $m$ values of the
pulsations. The bright ($m_{\rm v}=13.5$) DB GD358 fits these
criteria.  GD358 is the best studied DB pulsator, and the only DB with
existing, detailed asteroseismology (Winget et al. 1994 hereafter W94,
Bradley \& Winget 1994, Kepler et al. 2003 hereafter K03, Metcalfe et
al. 2003).

In cooperation with the Delaware Asteroseismic Research Center
(Provencal et al. 2005), we organized a WET run in May of 2006
(XCOV25) with GD358 as the prime target for Montgomery's light curve
fitting technique.  Our purpose was twofold: 1) obtain at least 5
hours of S/N$\approx$1000 photometry and 2) accurately identify the
frequencies, amplitudes and phases present in GD358's current
pulsation spectrum. The 2006 XCOV25 data set contains $\approx$436 hrs
of observations, with $\approx 29$ hrs of high S/N photometry.  While
GD358 is the best studied DB pulsator, the star's behavior is by no
means well understood.  The 2006 data set contains a great deal of
interesting information on GD358's pulsational behavior.  Before we
can proceed with detailed lightcurve fitting, we must thoroughly
understand the data set, examine GD358's pulsation spectrum, and
extract accurate frequency, amplitude, and phase information. In the
following, we will present an overview of the data set and reduction
procedures.  Our discussion will expand the existing
asteroseismological analysis of GD358, exploring the 2006 XCOV25
identified modes, combination frequencies, and multiplet structure. We
will compare our results with previous observations and examine the
complexity and evolution of GD358's pulsations over time. Our
investigation will lead us to consider connections between GD358's
pulsations, its convection zone, and a possible surface magnetic
field. The remarkable event that occurred in August 1996 (K03) will
play an important role in this discussion.  Finally, we present
implications for our understanding of GD358's physical properties and
suggest future investigations.

\section {Observations and Data Reduction}

GD358 (V777 Her), the brightest ($m_{\rm v}=13.7$) and best studied
helium atmosphere white dwarf pulsator, was the target of the 25th
Whole Earth Telescope (WET) run (XCOV25), the first sponsored by the
Delaware Asteroseismic Research Center (DARC).  The observations span
May 12 to June 16 2006, with the densest coverage between May 19 and
May 31. Nineteen observatories distributed around the globe
contributed a total of 88 individual runs (Table~1). The final XCOV25
light curve contains 436.1 hours of high speed photometry.

A goal of any WET run is to minimize data artifacts by obtaining as
uniform a data set as possible (Nather et al. 1990).  Early WET runs
(e.g. Winget et al. 1991) comprised mainly 3-channel blue-sensitive
photomultiplier tube (PMT) photometers that monitored the target
variable, a comparison star, and sky brightness. CCDs now bring
improved sensitivity and better sky measurements, but individual CCDs
have distinct effective bandpasses resulting in different measured
amplitudes from different observing sites. Recent WET runs (examples
include Kanaan et al. 2005, Vuckovic et al. 2006) comprise a mixture
of CCD and PMT observations, and XCOV25 is no exception.  CCDs were
employed at sixteen 
\linebreak
\LongTables
\begin{deluxetable*}{lllll}
\tablecolumns{5}
\tablewidth{0pc}
\tablecaption{Journal of 2006 XCOV25 Observations}
\tablehead{
\colhead{Run Name} &  \colhead{Telescope} & \colhead{Instrument}&  \colhead{Date}   & \colhead{Length} \\
\colhead{} & \colhead{}   & \colhead{}  &\colhead{}  & \colhead{(hrs)}
}
\startdata
konk20060512 & Konkoly 1.0m & CCD & 2006 May 12 &  6.8 \\
konk20060515 & Konkoly 1.0m & CCD& 2006 May 15 &  6.3 \\
mole20060515 & Moletai 1.65m & PMT & 2006 May 15 &  1.3 \\
mole20060517 & Moletai 1.65m & PMT & 2006 May 17 &  3.5 \\
konk20060517 & Konkoly 1.0m & CCD & 2006 May 17 &  5.3 \\
cuba20060517 &  Mt. Cuba 0.6m & CCD & 2006 May 18 &  5.4 \\
kpno20060518 &  KPNO 2.1m & CCD & 2006 May 18 &  7.0 \\
ctio20060518 &  CTIO 0.9m & CCD &2006 May 19 &  4.4 \\
kpno20060519 &  KPNO 2.1m & CCD& 2006 May 19 &  7.3 \\
hawa20060518 &   Hawaii 0.6m & CCD&  2006 May 19 &  2.2 \\
luln20060519 & Lulin 1.0m & CCD & 2006 May 19 &  5.0 \\
cuba20060519 & Mt. Cuba 0.6m & CCD & 2006 May 20 &  2.5 \\
ctio20060519 & CTIO 0.9m     & CCD & 2006 May 20 &  2.1 \\
kpno20060520 & KPNO 2.1m & CCD &2006 May 20 &  7.6 \\
hawa20060519 & Hawaii 0.6m & CCD & 2006 May 20 &  1.7 \\
ters20060520 & Peak Terskol 2.0m & CCD  & 2006 May 20 &  5.8 \\
cuba20060520 & Mt. Cuba 0.6m & CCD & 2006 May 21 &  7.1 \\
kpno20060521 & KPNO 2.1m & CCD & 2006 May 21 & 7.3\\
ctio20060520 & CTIO 0.9m & CCD & 2006 May 21 &  0.7 \\
hawa20060520 & Hawaii 0.6m & CCD & 2006 May 21 &  3.5 \\
ters20060521 & Peak Terskol 2.0m & CCD & 2006 May 21 &  6.0 \\
tueb20060521 & Tuebingen 0.8m & CCD & 2006 May 21 &  6.6 \\
lapa20060521 & La Palma WHT 4.2m & CCD & 2006 May 22 &  0.9 \\
cuba20060521 & Mt. Cuba 0.6m & CCD & 2006 May 22 &  4.4 \\
ctio20060521 & CTIO 0.9m & CCD & 2006 May 22 &  4.5 \\
kpno20060522 & KPNO 2.1m & CCD & 2006 May 22 &  1.0 \\
cuba20060521 & Mt. Cuba 0.6m & CCD & 2006 May 22 &  1.7 \\
hawa20060521 & Hawaii 0.6m & CCD & 2006 May 22 &  7.8 \\
luln20060522 & Lulin 1.0m & CCD & 2006 May 22 &  4.5 \\
konk20060522 & Konkoly 1.0m & CCD & 2006 May 22 &  4.2 \\
ters20060522 & Peak Terskol 2.0m & CCD & 2006 May 22 &  1.3 \\
mcdo20060523 & McDonald 2.1m & CCD & 2006 May 23 &  7.4 \\
kpno20060523 & KPNO 2.1m & CCD & 2006 May 23 & 2.9 \\
hawa20060522 & Hawaii 0.6m & CCD & 2006 May 23 &  6.6 \\
boao20060523 & BOAO 1.8m & CCD & 2006 May 23 &  7.1 \\
mcdo20060524 & McDonald 2.1m & CCD & 2006 May 24 &  7.2 \\
cuba20060523 & Mt. Cuba 0.6m & CCD & 2006 May 24 &  4.0 \\
hawa20060523 & Hawaii 0.6m & CCD & 2006 May 24 &  3.7 \\
boao20060524 & BOAO 1.8m & CCD & 2006 May 24 &  6.7 \\
ters20060524 & Peak Terskol 2.0m & CCD & 2006 May 24 &  6.1 \\
haut20060524 & OHP 1.93m & PMT & 2006 May 24 &  2.3 \\
mcdo20060525 & McDonald 2.1m & CCD & 2006 May 25 &  6.4 \\
hawa20060524 & Hawaii 0.6m & CCD& 2006 May 25 &  5.7 \\
ters20060525 & Peak Terskol 2.0m & CCD & 2006 May 25 &  7.1 \\
haut20060525 &  OHP 1.93m & PMT & 2006 May 25 &  4.2 \\
ctio20060525 & CTIO 0.9m & CCD & 2006 May 26 &  4.0 \\
hawa20060525 & Hawaii 0.6m & CCD & 2006 May 26 &  8.8 \\
haut20060526 & OHP 1.93m & PMT & 2006 May 26 &  4.0 \\
ters20060526 & Peak Terskol 2.0m & CCD & 2006 May 27 &  7.1 \\
ctio20060526 & CTIO 0.9m & CCD & 2006 May 27 &  4.3 \\
hawa20060526 & Hawaii 0.6m & CCD & 2006 May 27 &  9.1 \\
chin20060527 & BAO 2.16m & PMT & 2006 May 27 &  4.5 \\
haut20060527 &  OHP 1.93m & PMT & 2006 May 27 &  4.6 \\
cuba20060527 & Mt. Cuba 0.6m & CCD  & 2006 May 28 &  6.2 \\
lna20060528 & Itajuba 1.6m & CCD & 2006 May 28 &  3.1 \\
mcdo20060528 & McDonald 2.1m & CCD & 2006 May 28 &  0.7 \\
mcdo20060528b & McDonald 2.1m & CCD & 2006 May 28 &  7.2 \\
hawa20060527 & Hawaii 0.6m & CCD & 2006 May 28 &  7.8 \\
chin20060528 & BAO 2.16m & PMT & 2006 May 28 &  5.2 \\
haut20060528 &  OHP 1.93m & PMT & 2006 May 28 &  5.2 \\
vien20060528 & Vienna 1.0m & CCD & 2006 May 28 &  2.2 \\
cuba20060528 & Mt. Cuba 0.6m & CCD & 2006 May 29 &  3.4 \\
mcdo20060529 & McDonald 2.1m & CCD & 2006 May 29 &  8.2 \\
cuba20060528 & Mt. Cuba 0.6m & CCD & 2006 May 29 &  2.6 \\
hawa20060528 & Hawaii 0.6m & CCD & 2006 May 29 &  9.1 \\
ters20060529 & Peak Terskol 2.0m & CCD & 2006 May 29 &  6.0 \\
haut20060529&  OHP 1.93m & PMT & 2006 May 29 &  5.3 \\
cuba20060529 & Mt. Cuba 0.6m & CCD & 2006 May 30 &  2.7 \\
ters20060530 & Peak Terskol 2.0m & CCD & 2006 May 30 &  6.9 \\
haut20060530 & OHP 1.93m & PMT & 2006 May 30 &  4.2 \\
hawa20060530 & Hawaii 0.6m & CCD &  2006 May 31 &  8.7 \\
chin20060531 & BAO 2.16m & PMT & 2006 May 31 &  3.9 \\
ters20060531 & Peak Terskol 2.0m & CCD & 2006 May 31 &  7.0 \\
nord20060606 & NOT 2.7m & CCD & 2006 June  6 &  5.9 \\
tueb20060607 & Tuebingen 0.8m & CCD & 2006 June  7 &  4.2 \\
nord20060607 & NOT 2.7m & CCD & 2006 June  7 &  7.1 \\
tueb20060608 & Tuebingen 0.8m & CCD & 2006 June  8 &  4.7 \\
nord20060608 & NOT 2.7m & CCD & 2006 June  8 &  8.0 \\
tueb20060609 & Tuebingen 0.8m & CCD & 2006 June  9 &  5.4 \\
nord20060609 & NOT 2.7m & CCD & 2006 June  9 &  7.9 \\
tueb20060610 & Tuebingen 0.8m & CCD & 2006 June 10 &  5.2 \\
tueb20060611 & Tuebingen 0.8m & CCD & 2006 June 11 &  5.1 \\
tueb20060612 & Tuebingen 0.8m & CCD & 2006 June 12 &  4.5 \\
lapa20060613 & La Palma WHT 4.2m & CCD & 2006 June 13 & 1.8 \\
tueb20060613 & Tuebingen 0.8m & CCD & 2006 June 13 &  5.2 \\
lapa20060614 & La Palma WHT 4.2m & CCD & 2006 June 14 & 1.4 \\
lapa20060615 & La Palma WHT 4.2m & CCD & 2006 June 15 & 2.6 \\
lapa20060616 & La Palma WHT 4.2m & CCD & 2006 June 16 & 2.0 \\
\enddata
\end{deluxetable*}

\hspace*{-2.2em}
observatories, and 3-channel PMT photometers at
the remaining three sites.  We attempt to minimize bandpass issues by
using CCDs with similar detectors and equipping each CCD with a BG40
or S8612 filter to normalize wavelength response and reduce extinction
effects.  The bi-alkali photomultiplier tubes are blue sensitive, so
no filters are required. We also made every attempt to observe the
same comparison star at each site, but given plate scale differences,
that is not always possible.

We employ a 10 s contiguous integration time with the PMT photometers,
while the cycle time for the CCD observations, including exposure and
readout times, varies for each instrument.  To illustrate the
extremes, the Argos camera (Nather \& Mukadam 2004) at McDonald
Observatory uses a frame-transfer CCD and 5 s integrations with
negligible deadtime, while the camera on the CTIO 0.9m telescope
operated by the SMARTS consortium uses 10 second integration and a 25
second readtime for a total cycle time of $\approx$35 seconds.

Data reduction for the PMT observations follows the prescription
outlined in Nather et al. (1990) and W94. In most cases, a third
channel continuously monitored sky, allowing point by point sky
subtraction.  For two channel observations, the telescope is
occasionally moved off the target/comparison for sky observations. We
examined each light curve for photometric quality, and marked and
discarded ``bad'' points. ``Bad'' points are those dominated by
external effects such as cosmic rays or clouds.  We divided GD358's
light curve by the comparison star to remove first order extinction
and transparency effects. If necessary, we fit a low order polynomial
to the individual light curves to remove remaining low frequency
variations arising from differential color extinction.  We divided by
the mean light curve value and subtracted 1, resulting in a light
curve with amplitude variations as fractional intensity (mmi). The
unit is a linear representation of the fractional intensity of
modulation (1 mmi $\approx$1 mmag). We present our Fourier transforms
(FT) in units of modulation amplitude (1 mma = $1\times10^{-3}$ma).

XCOV25 marked an evolution of WET headquarters' role in CCD data
reduction. Standard procedure for a WET run is for observers to
transfer observations to headquarters for analysis at the end of each
night. In the past, CCD observers completed initial reductions (bias,
flat, and dark removal) at their individual sites, performed
preliminary aperture photometry, and transferred the result to WET
headquarters. For XCOV25, the majority of participants transferred
their raw images, enabling headquarters to funnel data through a
uniform reduction pipeline. The few sites unable to transfer images
nightly performed preliminary reductions on site using the same
procedures as those at headquarters, and sent their images at a later
date.

CCD data reduction follows the pipeline described by Kanaan et al.
(2002).  We corrected each image for bias and dark counts, and divided
by the flat-field. Aperture photometry with a range of aperture sizes
was performed on each image for the target and selected comparison
stars.  For each individual nightly run, we chose the combination of
aperture size and comparison star producing the highest signal/noise
as the final light curve. Each reduced CCD light curve consists of a
list of times and fractional intensities. As with the photometer
observations, we examined each light curve for photometric quality and
discarded ``bad'' points.

Finally, we combined the individual PMT and CCD light curves to
produce the complete light curve for XCOV25. This last step requires
detailed assessment of overlapping observations.  We make two
assumptions in this process: 1) our observational technique is not
sensitive to periods longer than a few hours, and 2) we assume GD358
oscillates about a mean light level. These assumptions allow us to
carefully identify and correct vertical offsets in overlapping
segments.
 
This data set does contain a significant fraction of overlapping
lightcurves. We experimented with the effects of overlapping data on
the FT by computing FTs with 1) all data included, 2) no overlapping
data, where we kept those data with higher signal to noise ratio, and
3) weighting the overlapping light curves by telescope aperture size.
We found no significant differences between the noise levels or FTs of
overlapping versus non-overlapping versus weighted data.
 
\begin{figure*}
  \centering{
\includegraphics[width=0.8\textwidth,angle=0]{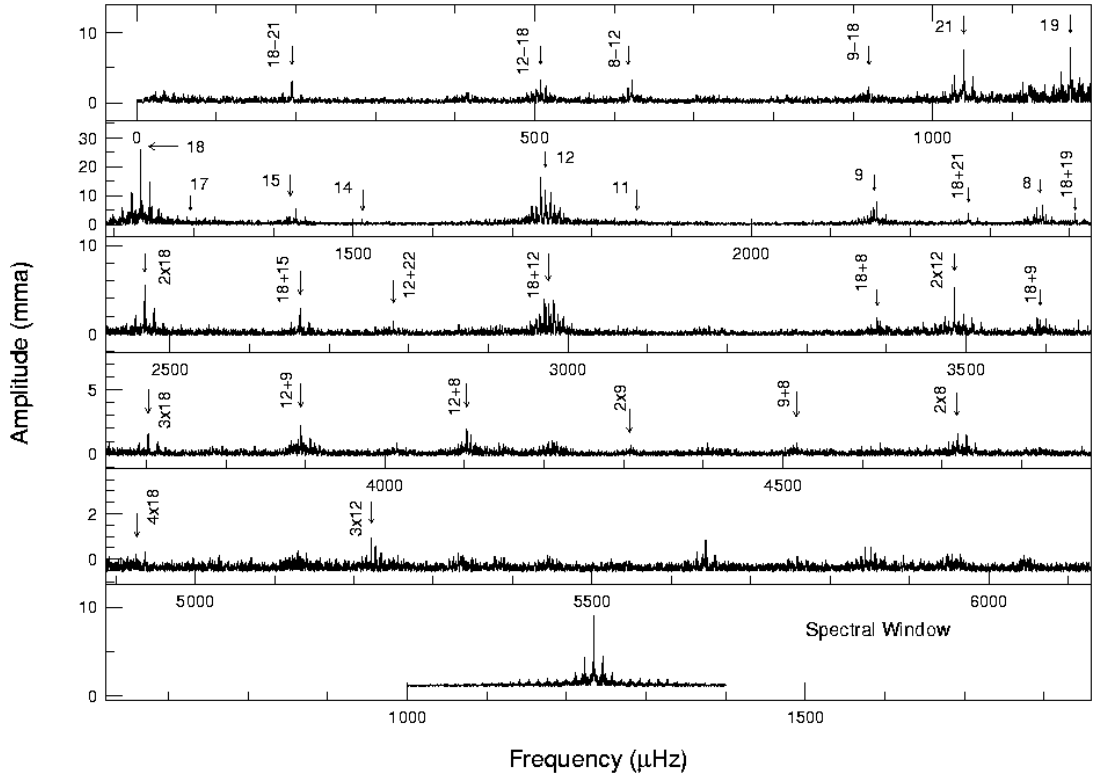}
}
\caption{Fourier Transform of the 2006 GD358 photometry observations 
  (note vertical scale in each panel). Arrows are labeled with $k$
  values for independent modes (single values) and first order
  combination frequencies. The unlabeled peaks are second and third
  order combinations.  The spectral window is plotted in the last
  panel.  Table~2 lists exact identifications.}
\end{figure*}

\begin{figure*}
  \centering{
 \includegraphics[width=0.8\textwidth,angle=0]{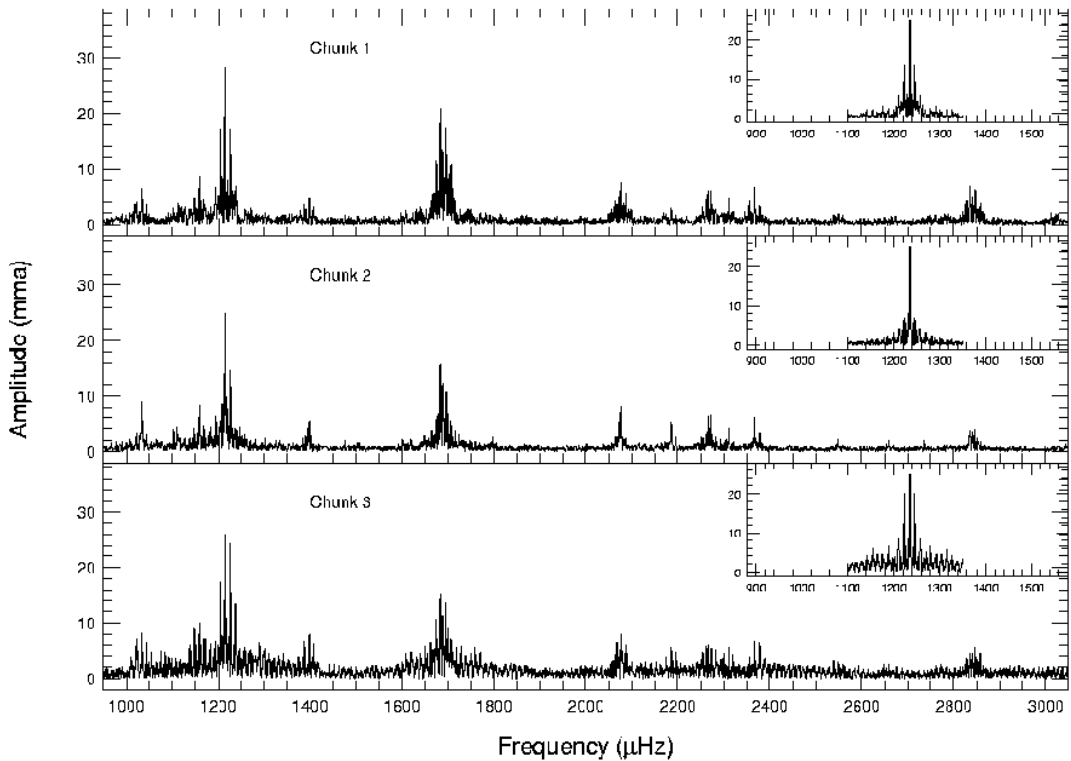}
 }
 \caption{FTs of the 2006 data set subdivided into three chunks of $\approx$180 hr 
   each. The changes in each FT can be explained as differences
   in the window structure/resolution for each chunk. GD358 was relatively 
   stable over this time frame.}
\end{figure*}

Despite the favorable weather enjoyed by many participating sites,
there are gaps within our light curve, especially near the beginning
and end of XCOV25 when fewer telescopes were on-line.  These gaps
produce spectral leakage in the amplitude spectrum, resulting in a
pattern of alias peaks associated with each physical mode that is not
of astrophysical origin. To quantify this, we sampled a single
sinusoid exactly as our original data and calculated its amplitude
spectrum.  The ``spectral window'' is the pattern produced by a single
frequency in our data set.  The Fourier transform and spectral window
of the final complete light curve are given in Figure~1.

\section{Frequency Identification}

\subsection{Stability}

Before we can look in detail at the 2006 XCOV25 FT, we must
investigate GD358's amplitude and frequency stability over the entire
timebase.  GD358 is known for large scale changes in amplitudes and
small but not insignificant frequency variations on a variety of
timescales (K03). Amplitude and/or frequency variations produce
artifacts in FTs, greatly complicating any analysis.  We divided the
data set into three chunks spanning $\approx$180 hrs and calculated
the FT of each chunk (Figure~2). For the dominant peak, the
frequencies are consistent to within measurement error and the
amplitudes remain stable to within 3$\sigma$.  The differences between
each FT are explained by variation in window structure and resolution
from chunk to chunk. We are confident that GD358 was fairly stable
over the length of XCOV25.

\subsection{The 2006 Fourier Transform}

\begin{figure}
\includegraphics[width=1.0\columnwidth,angle=0]{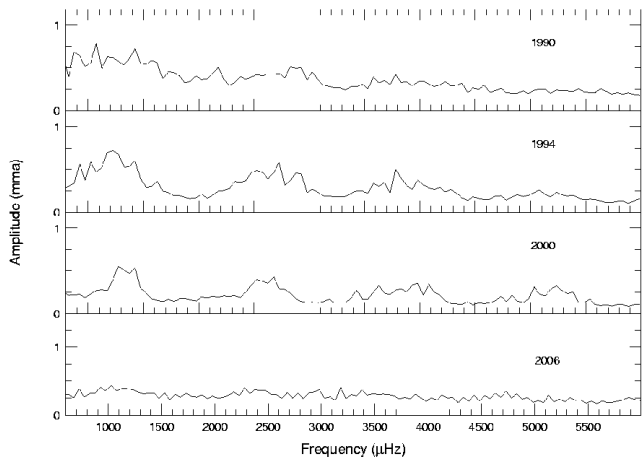}
 \caption{A comparison of the average noise from the 1990, 1994, 2000, and 2006
   WET runs. Each data set was prewhitened by the 50 largest amplitude
   frequencies. This is a conservative estimate of the noise in each
   data set.}
\end{figure}

To select the statistically significant peaks in the Fourier
transform, we adopt the criterion that a peak have an amplitude at
least 4.0 times greater than the average noise level in the given
frequency range. This leads to a 99.9\% probability that the peak is a
real signal present in the data, and is not due to noise (Scargle
1982, Breger et al. 1993, K03, for example). Here, ``noise'' is
defined as the average amplitude after prewhitening by the dominant
frequencies in the FT, and is frequency dependent.  This is a
conservative estimate, as it is impossible to ensure that all of the
``real'' frequencies are removed when calculating the noise level.
This is most certainly the case in GD358, where the FT contains myriad
combination frequencies. Figure~3 displays the average amplitude,
specified as the square root of the average power after prewhitening
by 50 frequencies, as a function of frequency, for 2006, and previous
WET runs on GD358 in 2000, 1994, and 1990.

The amplitude limit we select is very important in determining
``real''signals.  To confirm our uncertainty estimates, we performed
several Monte Carlo simulations using the routine provided in {\sl
  Period04}, software devoted to the statistical analysis of time
series photometry (Lenz \& Breger 2005). The Monte Carlo routine
generates a set of light curves using the original times, the fitted
frequencies and amplitudes, and added Gaussian noise. A least squares
fit is performed on each simulated light curve, with the distribution
of fit parameters giving the uncertainties.  Our Monte Carlo results
are consistent with our average amplitude noise estimates (Table~2),
confirming our use of the average amplitude after prewhitening.

\begin{deluxetable}{rrrrr}
\tablecolumns{4}
\tablewidth{0pc}
\tablecaption{Table of 2006 Independent Frequencies}
\tablehead{
\colhead {k (l=1)} &\colhead{2006 Frequency} &\colhead{Amp} & S/N & MC $\sigma_{\rm amp}$ \\
\colhead{} & \colhead{(\muHz)} & \colhead{(mma)} & \colhead{} & \colhead{(mma)}
}
\startdata
21 & $1039.076\pm0.002$ & $7.96\pm0.07$ & 27& 0.08\\
19 & $1173.042\pm0.002$ & $7.28\pm0.07$ & 25 & 0.08\\
18 & $1222.751\pm0.001$ & $5.0\pm0.07$ & 18 & 0.6 \\
18 & $1228.792\pm0.002$ & $5.7\pm0.07$ & 19 & 0.07\\
18 & $1234.124\pm0.001$& $24\pm0.07$ & 88 & 0.09 \\
18 & $1239.540\pm0.002$ & $4.7\pm0.07$ & 18 & 0.07 \\
18 & $1245.220\pm0.003$ & $4.7\pm0.07$ & 17 & 0.6 \\
17 & $1295.533\pm0.008$ & $1.4\pm0.08$ & 5 & 0.6 \\
15 & $1421.012\pm0.008$ & $2.8\pm0.08$ & 7 & 0.07\\
15 & $1423.942\pm0.008$ & $1.3\pm0.08$ & 6 & 0.07\\
15 & $1429.210\pm0.002$ & $5.6\pm0.07$ & 23 & 0.07\\
15 & $1434.784\pm0.008$ & $1.4\pm0.07$ & 4.4 & 0.06\\
15 & $1440.622\pm0.008$ & $1.4\pm0.07$ & 4.3 & 0.06\\
14 & $1512.010\pm0.007$ & $3.6\pm0.08$ & 8 &0.07 \\
12 & $1736.302\pm0.001$ & $17.0\pm0.07$ & 75 & 0.07\\
12 & $1737.962\pm0.007$ & $5.6\pm0.07$ & 8 & 0.08 \\
12 & $1741.665\pm0.001$ & $10.9\pm0.07$ & 50 & 0.07\\
12 & $1743.738\pm0.002$ & $5.6\pm0.07$ & 8 & 0.07 \\
12 & $1746.673\pm0.007$ & $1.8\pm0.08$ & 8 & 0.08 \\
12 & $1749.083\pm0.001$ & $12.9\pm0.07$ & 50 & 0.07 \\
11 & $1856.845\pm0.009$ & $1.4\pm0.08$ & 6.4 &0.08 \\
$9^{-1}$  & $2150.395\pm0.003$ & $4.2\pm0.07$ & 17  & 0.07 \\
$9^{0}$ &$2154.222\pm0.002$ & $5.5\pm0.07$ & 22 &0.07\\
$9^{+1}$ &$2158.071\pm0.002$ & $7.2\pm0.07$ & 29 & 0.07\\
$8^{-1}$ & $2359.064\pm0.002$ & $5.87\pm0.07$ & 22 &0.07 \\
$8^{0}$ & $2363.058\pm0.007$ & $1.82\pm0.08$ & 6 &0.07\\
$8^{+1}$ & $2366.523\pm0.002$ & $6.54\pm0.07$ & 23 &0.07\\
\enddata
\end{deluxetable}

For both Fourier analysis and multiple least squares fitting, we use
the program {\sl Period04}.  The basic method involves identifying the
largest amplitude resolved peak in the FT, subtracting that sinusoid
from the original light curve, recomputing the FT, examining the
residuals, and repeating the process. This technique is fraught with
peril, especially in multiperiodic stars, where it is possible for
overlapping spectral windows to conspire to produce alias amplitudes
larger than real signals. Elimination of this alias issue was the
driving motivation behind the development of the Whole Earth
Telescope, whose goal is to obtain nearly continuous coverage over a
long time baseline.  Our current data set on GD~358 does contain gaps,
but we have minimized the alias problem.

To illustrate the procedure we followed, let us examine the region of
dominant power at 1234 $\muHz$ (Figure~4). Comparison of this region
with the spectral window demonstrates that most of the signal is
concentrated at 1234 $\muHz$. We fit the data with a sine wave to
determine frequency, amplitude, and phase, and subtract the result
from the original light curve.  The second panel of Figure~4 shows the
FT prewhitened by this frequency.  Careful examination reveals two
residual peaks (arrows) that are clearly not components of the
spectral window.  We next subtract a simultaneous fit of the 1234
$\muHz$ frequency and these two additional frequencies, with the
results displayed in panel 3 of Figure~4. At this point, we must
proceed with extreme caution.  The remaining peaks, which correspond
closely with aliases in the spectral window, are significant and
cannot be ignored. We are faced with two possibilities: 1) these peaks
represent real signal, and 2) amplitude modulation is present.  While
we have ruled out large modulations in frequency/amplitude during
XCOV25, small scale amplitude modulation may be intrinsic to GD358, or
artificially present in the data set. An examination of data from
individual sites reveals that although the frequencies from each site
are the same within the statistical uncertainties, two observatories,
one using a CCD and one a PMT, report consistently lower amplitudes,
probably due to beating effects.  Removing the suspect observations
from the data set does not alter the prewhitening results for the 1234
$\muHz$ region.  As an additional test, we closely examined the FTs
from Figure~2.  Prewhitening the 1234 \muHz\ regions of each chunk
consistently produces the same 5 peaks.  Finally, we note that one of
these peaks appears in combination with other modes (Table~3).
Therefore, we believe our first possibility is most probable, all five
peaks are real and are listed in Table~2.

\begin{figure}
\includegraphics[width=1.0\columnwidth,angle=0]{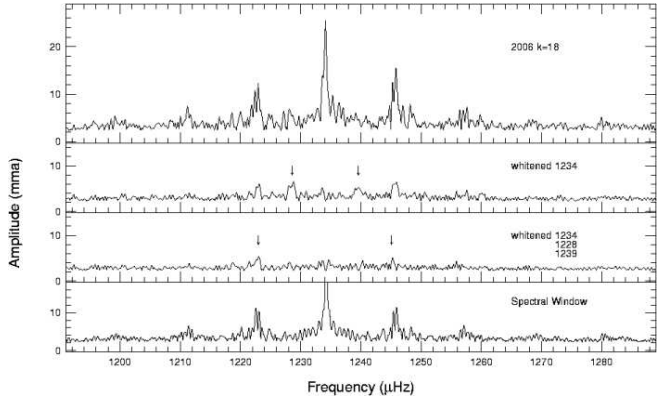}
 \caption{Demonstration of prewhitening using the dominant 1234 \muHz\
   mode in the 2006 FT. We begin with the removal of the largest
   amplitude resolved peak, a careful comparison of the residuals with
   the spectral window (last panel), and the subsequent removal of
   additional peaks. }
\end{figure}

\begin{figure}
\includegraphics[width=1.00\columnwidth,angle=0]{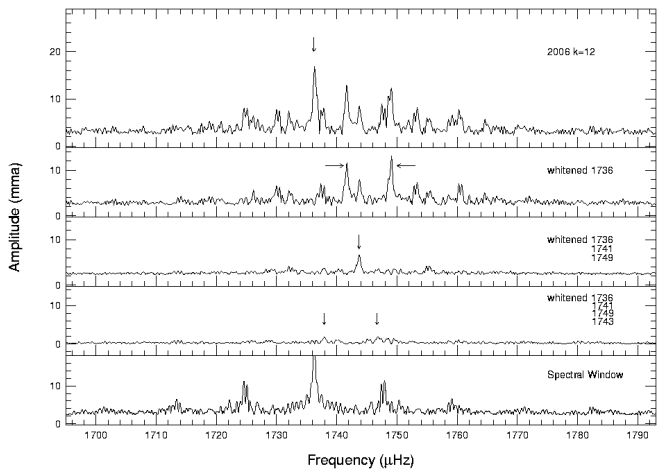}
\caption{Prewhitening of the complex $k=12$ mode. We find three peaks 
  over 10 mma at 1736, 1741 and 1749 \muHz, and numerous additional
  peaks (Table~2).  This mode could contain as many as 9 components.}
\end{figure}

\begin{figure}
\includegraphics[width=1.00\columnwidth,angle=0]{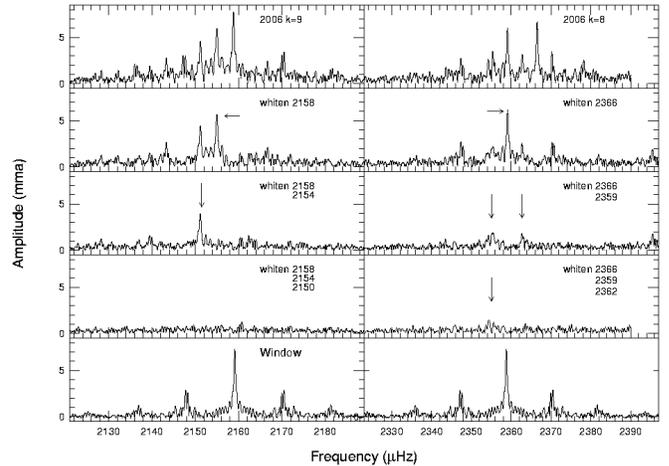}
\caption{Prewhitening of the $k=9$ and $8$ modes. These modes are 
  triplets consistent with previous reports of W94 and K03.}
\end{figure}

We employed this procedure to identify frequencies satisfying our
criteria of amplitudes $4\sigma$ above the average noise. The large
amplitude power at 1740 \muHz\ is particularly complex. Unlike our
example at 1234 \muHz, this mode contains 3 peaks with amplitudes over
10 mma and contains as many as 9 frequencies (Figure~5). Figure~6
demonstrates the prewhitening results for modes at 2154 and 2362
\muHz.

Our final identifications result from a simultaneous nonlinear least
squares fit of 130 frequencies, amplitudes, and phases.  We adopt the
$l$=1 mode identifications for GD358 established in W94. We emphasize,
however, that $k$, representing the number of nodes in the radial
component of displacement from the center to the stellar surface,
cannot be observationally determined and may not correspond precisely
to the values given here.  Tables~2 lists 27 identified independent
frequencies. Table~3 presents significant combination frequencies. In
the interest of space, we do not list all of the combination
frequencies.

\section{GD358 in 2006}

\subsection{Independent Modes}

\begin{figure}
\includegraphics[width=1.00\columnwidth,angle=0]{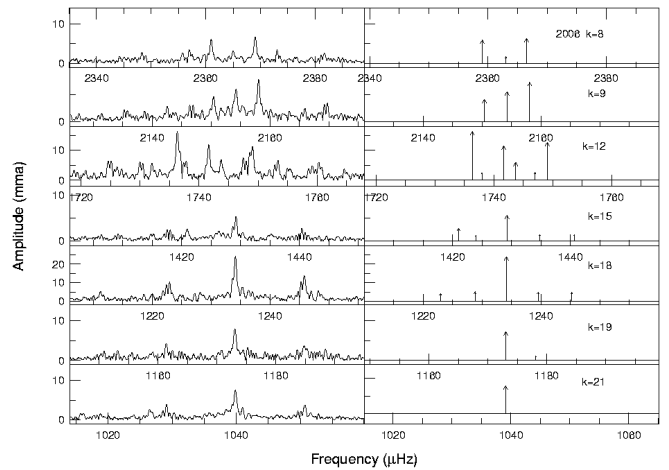}
 \caption{A ``snapshot'' of the 2006 largest amplitude modes and their 
   multiplet structure.  The left panels plot the observed FT, and the
   right panels presents the prewhitening results.  The 1$\sigma$
   frequency errors are $\approx$0.001 \muHz. Each panel spans 50
   \muHz. Note the changes in y-scale.}
\end{figure}

The 2006 XCOV25 data set illustrates GD358's continuing tendency for
changing the distribution of amplitudes among its excited modes.
Using the $k$ identifications established in W94, we detect power at
$k$=21, 19, 18, 17, 15, 14, 12, 11, 9, and 8.  The principal frequency
is at 1234 \muHz\ ($k=18$) with an amplitude of 24 mma. The previously
dominant $k=17$ and 15 modes (K03) are greatly diminished in amplitude
(1.4 and 5.6 mma respectively), and we do not detect the $k=16$ or
$k=13$ modes reported by W94 and K03.  We cannot identify the
suspected $l=2$ mode at 1255.4 \muHz\ or $k=7$ at 2675.5 \muHz\ noted
by K03).  K03 suggest that $k=7$ may have been excited to visibility
via resonant coupling with $k=17$ and 16.  Since neither has
significant amplitude in 2006 (Figure~1), it logically follows that this
mode would not be detected. Perhaps the greatest surprise from XCOV25
is the appearance of prominent power at the predicted value for
$k=12$, a region of the FT previously devoid of significant peaks.

\clearpage

 \begin{deluxetable*}{rrrrrrr}
 \tablecolumns{5}
 \tablewidth{0pc}
 \tablecaption{2006 Combination Frequencies}
 \tablehead{
 \colhead{Frequency} &\colhead{Combination(s)} &\colhead{$f_{\rm obs}-f_{\rm calc}$} &\colhead{Amp} &\colhead(S/N) & $R_c$\\
 \colhead{(\muHz)} &\colhead{k} &\colhead{$(\muHz)$} &\colhead{(mma)} &\colhead{}
 }

 \startdata
 195.085  & ${1234-21}$ & 0.04 & 2.7 & 7.7 & - \\
 507.523  & ${1741.7-1234}$ & -0.02 & 3.0 & 8.2 & - \\
 622.798  & ${2358.1-1736.3}$ & -1.0 & 2.8 & 7.5 & -\\
 920.039  & $\rm{9^{0}-1234}$ & -0.06 & 1.9 & 4.5&  - \\
 2078.187 & $\rm{2x1039.1}$ & -0.1 & 0.74 & 3.2 & 2.4\\
 2273.244 & $\rm{1234+21}$& 0.05 & 4.2 & 10.8 & 9.6\\
 2407.205 & $\rm{1234+19}$& 0.07 & 3.8 & 9.7 & 9.1\\
 
 2462.989 & $\rm{1234+1228.8}$ & 0.07 & 3.7 & 8.2 & 11.4\\
 2468.282 & $\rm{2x1234}$ & 0.034 & 5.1 & 13.1 & 1.8 \\
 2479.358 & $\rm{1234+1245.2}$ & 0.014 & 1.5 & 4.1 & 5.6 \\
 2663.369 & $\rm{1234+1429.2}$ & 0.04 & 2.9 & 7.3 & 9.4\\
 2780.786 & $\rm{1741.7+1039.1}$ & 0.014 & 1.4 & 4.0 & 7.1 \\

 2909.416 & $\rm{1736.3+19}$ & 0.1 & 1.0 & 2.6 & 3.5 \\

 2964.917 & $\rm{1228.8+1736.3}$ & -0.18 & 1.1 & 2.8 & 5.0 \\
      & $\rm{1222.8+1741.7}$ & -0.46 & 1.1 & 2.8 \\
 2970.400 & $\rm{1234+1736.3}$& 0.025& 3.0 & 7.6 & 3.1\\
          & $\rm{1228.8+1741.7}$& -0.057 &  & \\
 2972.085 & $\rm{1234+1737.9}$& 0.036 & 2.82 & 7.6 & 8.8\\
          & $\rm{1222.8+1749.1}$& -0.22 & & \\
 2975.814 & $\rm{1234+1741.7}$ & -0.024 & 3.4 & 8.7 &5.5\\
          & $\rm{1239.5+1736.3}$ & -0.039 & & \\
 2977.885 & $\rm{1228.8+1749.1}$ &  0.01 & 1.7 & 4.4 & 11.9 \\
          & $\rm{1234+1743.8}$ & 0.02 & \\
 2981.032 & $\rm{1239.5+1741.7}$ & 0.17 & 1.3 & 4.0 & 11.2\\
 2981.947 & $\rm{1245.2+1736.3}$ & -0.45 & 3.5 & 7.9 & 20.6\\
 2983.266 & $\rm{1234+1749.1}$ & 0.057 & 3.1 & 7.9 & 4.66\\
 2988.643 & $\rm{1239.5+1749.1}$ & -0.42 & 1.1 & 3.9 & 9.36\\

 3388.400 & $\rm{1234+9^{0}}$ & 0.052 & 1.5 & 4.0 & 4.75 \\
 3392.185 & $\rm{1234+2150.4}$ & -0.013 & 1.7 & 4.4 & 7.35\\
 3472.552 & $\rm{2x1736.3}$ & -0.052 & 1.5 & 4.8 & 1.2 \\
 3477.939 & $\rm{1736.3+1741.7}$ & 0.06 & 1.7 & 4.5 & 4.0\\
 3485.387 & $\rm{1741.7+1743.7}$ & -0.02 & 5.7 & 16.7 & 11.6\\
      & $\rm{1736.3+1749.1}$ & 0.003 & 5.7 & 16.8 \\
 3490.720 & $\rm{1741.7+1749.1}$& 0.03 & 1.3 & 4.0 \\

 3593.136 & $\rm{1234+2359.1}$ & -0.041 & 1.6 & 4.2 & 4.8\\
 3600.577 & $\rm{1234+2366.5}$ & -0.071 & 1.7 & 4.5 & 4.5\\
 3702.443 & $\rm{3x1234}$ & 0.071 & 2.0 & 5.3 \\

 3890.607 & $\rm{1736.3+9^{0}}$ & 0.08 & 1.5 & 4.8 & 7.0\\
 3894.392 & $\rm{1736.3+2158.1}$ & 0.016 & 2.7 & 7.7 & 6.8\\
 3907.140 & $\rm{1749.083+2158.1}$ & -0.014 & 1.6 & 5 & 7.7\\
 4095.326 & $\rm{1736.302+2359.1}$ & -0.03 & 1.2 & 4.0 &5.2 \\
 4102.850 & $\rm{1736.302+2366.5}$ & 0.024 & 2.2 & 7.3 &4.0\\
 4108.130 & $\rm{1749.083+2359.1}$ & -0.06 & 1.7 & 5.6 & 9.8 \\
 4209.941 & $\rm{2\times1234+1741.7}$ & 0.028 & 1.3 & 4.8 \\
 4308.536 & $\rm{2x9^{0}}$ & -0.092 & 1.0 & 4.6 & 11.7\\
 4517.044 & $\rm{8^{0}+9^{0}}$ & -0.24 & 1.1 & 4.7& 48.5 \\
          & $\rm{2359.1+2158.1}$ & 0.13 & \\
          & $\rm{2150.4+2366.5}$ & -0.08 & \\
 4719.483 & $\rm{2x2359.064}$ & -1.4 & 2.1 & 7.2 & 13.2\\
 4718.268 & $\rm{2x2359.064}$ & -0.14 & 1.1 & 4.7 \\
 5221.734 & $\rm{1736.3+1741.7+1743.7}$ & 0.034 & 1.3 & 4.2 \\
         & $\rm{2x1736.3+1749.1}$ & -0.049 & \\
 5643.452 & $\rm{1736.3+1749.1+2158.1}$ & 0.05 & 1.2 & 4.1 \\
 \enddata
 \end{deluxetable*}

Figure~7 presents a ``snapshot'' of multiplet structure in the XCOV25
FT.  The only modes exhibiting structure with splittings in agreement
with previous observations are $k=9$ and $8$. We find an average
multiplet splittings of 3.83 and 3.75 \muHz, respectively. The
dominant $k=18$ is a quintuplet with an average splitting of 5.6
\muHz. The 5.6 \muHz\ splitting also appears in the $k=15$ and $k=12$
modes.  The $k=15$ mode contains 5 peaks, only 3 of which are split by
5.6 \muHz. The $k=12$ mode could contain as many as 9 components, if
we relax our 4.0 $\sigma$ detection criteria slightly. A possible
interpretation for $k=12$ is two, perhaps three, overlapping triplets,
each with $\approx5.6$ \muHz\ splitting. An $l$=2 mode is predicted at
$\approx1745$\muHz, so a second possible interpretation is an overlap
of $l$=1 and $l$=2 modes.  We point out that the other high $k$ modes
(17, 16, 14, and 13) reported by W94 to have frequency splittings of
$\approx6$ \muHz\ are either not detected here, or do not have
sufficient amplitude to investigate multiplet structure.

Our goal for XCOV25 is the identification/confirmation of $l$ and $m$
values for modes in GD358's 2006 light curve. This information is
required to fit our high signal to noise lightcurve using Montgomery's
technique.  However, the 2006 multiplet structure proves to be
puzzling. In the limit of slow rotation, we expect each mode to be
split into 2$l$+1 components.  Except for $k=9$ and $k$=8, we do not
find the distinct triplets attributable to $l$=1. Indeed, $k$=18 is a
quintuplet at first glance suggesting $l$=2, and $k$=15 and 12 are
both complex but lacking equal splittings.  We cannot confirm the $l$
or $m$ designations of 
\linebreak
 \begin{deluxetable}{ccc}
 \tablecolumns{3}
 \tablewidth{0pc}
 \tablecaption{Mean Period Spacing}
 \tablehead{
 \colhead{Year} &\colhead {Time} &\colhead{$\Delta P_{av}$}\\
 \colhead{} &\colhead{BJED} &\colhead{(s)}
 }
 \startdata
 1990 & 2440831.772 & $38.60\pm0.3$  \\
 1991 & 2448356.716 & $38.39\pm0.3$ \\
 1994 & 2449474.998 & $38.67\pm0.3$ \\
 2000 & 2451702.402 & $38.77\pm0.3$ \\
 2006 & 2453868.311 & $38.77\pm0.3$ \\
 \enddata
 \end{deluxetable}

\hspace*{-1.1em}
W94 and K03 based on multiplet structure alone.  Fortunately, the
identification of the spherical degree of GD358's pulsation modes as
$l$=1 does not rely solely on multiplet structure.  Table~4 presents
the mean period spacing for XCOV25, as well as WET runs from 1990,
1991, 1994 and 2000.  The average period spacing, which is dependent
on stellar mass, is 38.6 seconds. If GD358's modes are consecutive
$l$=1, the period spacing corresponds to a stellar mass of
$\approx0.6M_o$, in agreement with spectroscopic results (Beauchamp et
al. 1999).  If they are consecutive $l$=2 modes, the derived stellar
mass becomes $\approx0.2M_o$, making GD358 one of the lowest mass
field white dwarfs known and incompatible with the spectroscopic log
$g$. In addition, the distance of $42\pm3$pc derived from GD358 models
assuming $l$=1 is consistent with the trigonometric distance of
$36\pm4$pc.  The distance derived derived from models assuming $l$=2
is $\approx$75 pc (Bradley \& Winget 1994).  Is it possible that we
have an amalgam of $l$=1 and 2 modes?  $l$=2 modes are predicted to be
quintuplets, fitting with our results for $k$=18.  However, we expect
the ratio of rotational splittings between $l$=1 and $l$=2 to be
${R_{1,2}=0.60}$ (Winget et al. 1991). Assuming $k$=8 and 9 are $l$=1
as indicated by their triplet structure, we expect an $l$=2 splitting
of 6.3 \muHz, much larger than the observed 5.6 \muHz\ splitting. We
have no easy explanation for the 5.6 \muHz\ splitting, and hesitate to
identify $k$=18 as an $l$=2 mode solely on the basis of a quintuplet.
The $k$=12 and $k$=15 modes contain the mysterious 5.6 \muHz\
splitting, but the splittings are unequally spaced.  The ``best fit''
pulsation models do predict $l$=2 modes at 1250 \muHz\ (near $k$=18)
and 1745 \muHz\ (near $k$=12) (Metcalfe et al. 2003).  The 1250 \muHz\
mode is not detected in this data set. Some of the complex peaks at
$k$=12 could correspond to an underlying $l$=2 mode, but the
splittings do not support this conclusion.  Finally, optical and UV
radial velocity variations have been used to determine the $l$ values
for several modes in GD358, including $k$=18.  Kotak et al.  (2003)
identified radial velocities corresponding to $k$=18, 17, 15, 9, and 8
in their time-resolved optical spectra.  They determined that these
modes all share the same $l$ value, which is probably $l$=1.
Castanheira et al. (2005) use HST UV time resolved spectroscopy to
determine that $k$=9 and 8 are best explained as $l$=1.

While we are fairly certain that the majority of GD358's modes are
$l$=1, except for $k$=9 and 8, we cannot directly provide $m$
identifications for the observed multiplet components or confirm the
$m$ identifications of W94 and K03.  For this data set, we must resort
to indirect means.  In the next section, we examine the combination
frequencies, and, limiting ourselves to $l$=1, explore their potential
for pinning down the multiplet $m$ components in GD358.

\vspace*{2em}

\subsection {Combination Frequencies}

GD358's 2006 FT contains a rich distribution of combination
frequencies, from differences to simple harmonics to fourth order
combinations.  Combinations peaks are typically observed in the FTs of
large amplitude pulsators (Dolez et al. 2006, Yeates et al. 2005,
Thompson et al. 2003, Handler et al. 2002, W94).  Their frequencies
correspond to integer multiples of a single frequency (harmonics),
combinations (both sums and differences) of different components of a
given multiplet, or combinations (again both sums and differences) of
components of different modes (cross combinations). The general
consensus on the origin of these combinations dictates that they are
not independent modes, but are the result of nonlinear mixing induced
by the convection zone in the outer layers of the star (Brickhill
1992, Wu 2001, Ising \& Koester 2001). The convection zone varies its
depth in response to the pulsations, distorting and delaying the
original sinusoidal variations, and redistributing some power into
combination frequencies.  While combinations themselves do not provide
additional direct information about the interior structure of the
star, they are sensitive to mode geometry, making them potential tools
for identifying $l$ and/or $m$ values and orientation for each mode.

Wu (2001) provides a theoretical overview of combinations and their
interpretation.  Handler et al. (2002) and Yeates et al. (2005) apply
Wu's approach to several ZZ Ceti (DA) stars.  Yeates et al. (2005)
focus on the dimensionless ratio $R_c$ first introduced by van
Kerkwijk et al. (2000). The theoretical value of $R_c$ is given as
$R_c=\it{F(\omega_i,\omega_j,\tau_{c_0},2\beta+\gamma)}\it{G(l,m,\theta)}$.
The first term, $\it{F}$, includes the frequencies of the parent modes
($\omega_i, \omega_j$), as well as properties inherent to the star,
such as the depth of the stellar convection zone at equilibrium
($\tau_{c_0}$) and the sensitivity of the convection zone to changes
in temperature ($\beta, \gamma$).  The second term, $\it{G}$, includes
geometric factors such as the $l$ and $m$ values of the parent modes,
and the inclination angle of the pulsation axis. Both Wu (2001) and
Yeates et al. (2005) present theoretical predictions for various
combinations of $l$ and $m$ values.

The observed value of $R_c$ is defined as the ratio between the
observed amplitude of a combination and the product of the observed
amplitudes of its parent modes, including a correction factor
incorporating the bolometric correction and an estimate for the
convective exponent. The convective exponent appears in theoretical
estimations of the thermal response timescale of the convection zone.
From Montgomery (2008), we estimate the bolometric flux correction for
GD358 to be 0.35 and the convective exponent to be 25. For harmonic
combinations, this produces a corrective factor of 0.22. Cross
combinations include an additional factor of 2, resulting in a
correction of 0.44. With this formalism, and limiting ourselves to the
previous identification of the principal modes as $l$=1, and with the
smorgasbord of combination frequencies present in GD358, it should be
possible to determine $m$ values of the parent modes, to provide
limits on the inclination angle, and to further study convection in
white dwarfs.  However, interpretation of the combination frequencies
is not as straightforward as we would hope.

The analysis of $R_c$ for harmonic combinations is simplified by the
fact that we are certain that the combination contains a single
parent, and so single values for $l$ and $m$.  In the 2006 data set,
we detect the dominant 1234 \muHz\ mode's 1st, 2nd, and 3rd harmonics,
and place upper amplitude limits of $\approx0.3$ mma on higher orders.
The first harmonic, at 2468.282 \muHz\ with an amplitude of 5.1 mma,
is the 12th highest amplitude peak in the FT and the second highest
amplitude combination frequency observed.

The second largest mode and most complicated multiplet in the 2006 FT
is $k$=12. Focusing on its three large amplitude components (Table~3),
we find the 1st harmonic of 1736 \muHz, near 3472 \muHz\ with an
amplitude of 1.5 mma. Interestingly, the largest peak near 3472 \muHz\ 
is actually a sum of $k$=12 components (see Table~3), not a simple
harmonic, a behavior distinctly different from $k$=18. We place an
upper limit of 0.5 mma for a 1st harmonic of 1741.67 \muHz, 0.6 mma
for 1749.08 \muHz, and upper limits of 0.3 mma for all higher orders
for all other components of $k$=12.

For the remaining modes in Table~2, we detect the first harmonic of
1039 \muHz\ ($k$=21) with an amplitude of 0.7 mma. For $k$=19, 17, 15,
14, and 11, we place upper limits of 0.5 mma for their first
harmonics. The 2154 \muHz\ peak, identified as $k$=9, $m$=0
($k=9^{0}$) in W94, has a first harmonic with an amplitude of 1.0 mma.
It is surprising that we do not detect a harmonic of the larger
amplitude $k=9^{+1}$ component.  We place upper limits of 0.4 mma for
harmonics of the $k=9^{\pm1}$ components. Finally, the 2359 \muHz\ 
component of $k=8$, identified as $k=8^{-1}$ in W94, has a first
harmonic at 4719.483 \muHz\ (2.1 mma).  We place upper limits of 0.4
mma for harmonics of the $k=8^0$ and $k=8^{+1}$ components. We note
again that the larger amplitude $k=8^{+1}$ does not have a significant
harmonic.

\begin{figure}
\includegraphics[width=1.00\columnwidth,angle=0]{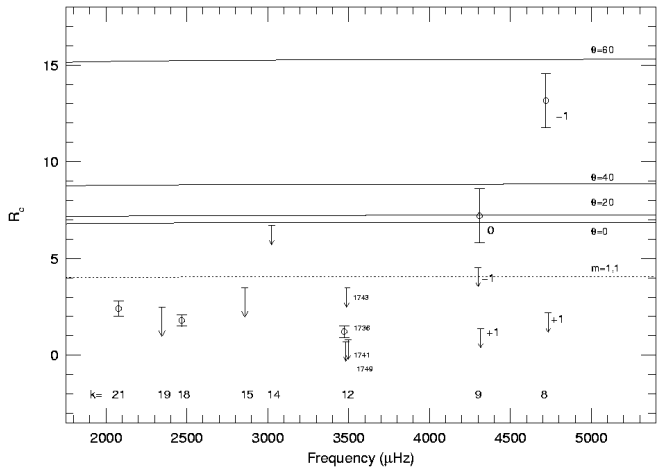}
 \caption{Observed $R_c$ values (ratio of amplitude of the combination
   to the product of the parent amplitudes) for detected
   harmonics. The open circles denote detected harmonics with estimated errors, 
   and the vectors plot upper limits on other harmonics. The solid lines
   represent theoretical values for the harmonics of $m=0$ modes for
   inclinations of 0, 20, 40 and 60 degrees. The dotted line gives
   the theoretical value for the harmonics of $m=1$ and $m=-1$ modes, which do not 
   depend on inclination. We rely on previous identifications of $m$ values for
   $k=9$ and $8$.}
\end{figure}

We begin by plotting the observed values of $R_c$ for first order
harmonics in Figure~8. If no harmonic is detected, we present upper
limits.  Since we do not have $m$ identifications for most modes but
we know that $m$ must be the same for harmonic parents, we plot
theoretical predictions for $l=1$, $m=0,0$ combinations for
inclinations of 0, 20, 40, and 60 degrees (solid lines) and $l=1$,
$m=1,1$ and $m=-1,-1$ (dotted line).  $R_c$ values for $m=1,1$ and
$m=-1,-1$ combinations do not depend on inclination.  Surprisingly,
the observed $R_c$ values and upper limits for the high $k$ modes are
lower than the theoretical predictions. We could argue that, under the
theoretical assumptions of Wu (2001), the detected high $k$ multiplet
components are not consistent with $m=0$.

From the $k=9$ harmonics, we find behavior closer to theoretical
predictions. The $k=9^0$ harmonic is consistent with an inclination of
between 0 and 40 degrees, while the $k=9^{\pm1}$ components are in the
direction of theoretical predictions. However, for $k$=8, the
$k=8^{-1}$ harmonic has a larger value of $R_c$ than the $k=8^{+1}$
upper limit, something that is difficult to explain using current
theory and simple viewing arguments.  The upper limit of $R_c$ for
$k=8^{0}$ is 29, and is not shown in Figure~8.

The most eye-catching example of cross combination peaks is found near
3000 \muHz\ (Figure~1), corresponding to linear combinations between
the two largest multiplets, $k$=18 and $k$=12. Our resolution and
sensitivity are sufficient to reveal 9 combinations, with amplitudes
ranging from 3.4 mma to 1.1 mma. In several cases, multiple parent
identifications are possible for each mode. For the following
discussion, we include identifications for detected combinations
involving at least one large amplitude parent, with most including the
1234 \muHz\ mode in combination with components of $k$=12 (Table~3).
The 1234 \muHz\ mode also produces large amplitude combinations (4.2
and 3.8 mma) with the 1039 ($k$=21) and 1173 ($k$=19) \muHz\ modes.

\begin{figure}
\includegraphics[width=1.0\columnwidth,angle=0]{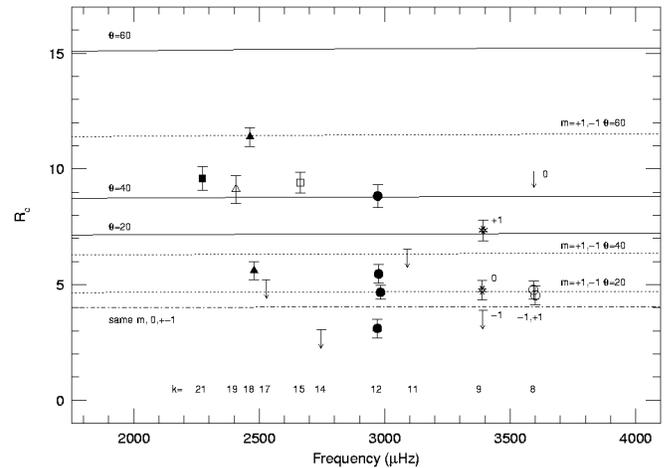}
 \caption{Observed $R_c$ values (ratio of amplitude of the combination 
   to the product of the parent amplitudes) for cross combinations
   including 1234 \muHz\ as one parent mode. The $k$ value of the
   second parent is given along the bottom of the figure, and
   different point types indicate combinations of 1234 \muHz\ with
   different $k$s. The solid lines represent theoretical predictions
   for $m=0,0$ combinations with inclinations of 20, 40 and 60
   degrees, the dotted lines are for $m=+1,-1$ combinations with the
   same inclinations, and the dot/dash line represents $m=1,1$ and
   $m=-1,-1$, and $m=0,\pm1$ combinations. Error bars are $1\sigma$
   formal errors. }
\end{figure}

Figure~9 presents observed $R_c$ values and upper limits for first
order cross combinations including the dominant 1234 \muHz\ mode as
one parent.  We include the same theoretical predictions as in Figure~8,
adding predictions for $l$=1, $m=+1,-1$ combinations for inclinations
of 20, 40, and 60 degrees (dotted lines), and $m=0,\pm1$ combinations
(dot/dash line).  The $R_c$ values for combinations of 1234 \muHz\ 
with the high $k$ modes 21, 19, 15, and one component of $k=12$ (1739
\muHz) are consistent with either $m=0,0$ combinations with an
inclination of roughly 40 degrees or $m=+1,-1$ combinations with an
inclination closer to 50 degrees. Recalling that the harmonic
combinations (Figure~8) argue that none of the high $k$ modes are $m=0$,
the second choice of $m=+1,-1$ combinations seems more likely.  This
further implies that 1234 \muHz\ is either $m=+1$ or $m=-1$, and the
other modes are all the remaining value.

The relative values of $R_c$ among the 3000 \muHz\ combinations
($k$=18+12), while independent of most model parameters, do depend on
inclination, and should reflect their respective projection in our
line of sight.  We should find $R_c$ values corresponding to
combinations of $m=0,\pm1$, $m=0,0$, $m=-1,+1$, and $m$=same. However,
even incorporating our tenuous identification of 1234 \muHz\ as
$m$=-1, given the number of multiplet components and large
uncertainties in the inclination angle, our analysis based on Wu
(2001) predictions produces no clear identification.

Our analysis of GD358's combination frequencies is complex.  We find a
very tenuous identification for the 1234 \muHz\ mode of $m=\pm1$,
requiring the remaining high $k$ modes' largest component to be the
remaining value. We find no clear $m$ identifications for the
components of $k$=12.  This mode is clearly more complex than the
simple multiplet structure expected in the limit of slow rotation.

There are multiple reasons why GD358's combination frequencies defy
simple interpretation. First, this star simply has too many principal
modes simultaneously excited to large amplitude. The cross-talk
between them is large,and the perturbative treatment of Wu (2001) may
not be valid. Second, both Wu (2001) and Yeates et al. (2005) assume
the intrinsic amplitudes are the same for every component in a
multiplet, and this may simply not be the case for GD358. Third, the
theoretical value of $R_c$ relies on correct identification of $l$ and
$m$, although in simpler cases, as in Yeates et al. (2005), it can be
used as a mode identification technique. While we are fairly certain
that GD358's pulsations are $l$=1, the assumption of a classical
triplet split by rotation is not valid for GD358's high $k$ modes.
Finally, we raise a possibility suggested by the $k$=9 and 8
components and their harmonics: to first order, the $m$=0 components
sample the radial direction, and the $m=\pm1$ components sample
azimuthal direction.  Components of the same mode have the same
inclination, so cancellation effects should be the same.  The $k$=9
and 8 multiplet amplitudes and presence/absence of harmonics argue
that something is interrupting the azimuthal symmetry of GD358.  It is
possible that the pulsations do not all share the same inclination, in
which case they would not combine as expected by Wu (2001). The
oblique pulsator model is well known, especially in the case of roAp
stars. For example, Bigot \& Dziembowski (2002) show that the main
mode of HR 3831, a pulsating roAp star, departs from alignment with
the magnetic axis.  GD358's combinations are trying to tell us
something important about the pulsation geometry, but we don't yet
understand how to interpret it.

While the general frequencies of the high $k$ multiplets agree with
the identifications of W94 and K03, the 2006 multiplets themselves are
complex, and cannot be explained by simply invoking rotation using
standard theory and simple geometric viewing arguments. Except for the
$k$=9 and $k$=8 modes, we do not find triplet structure as reported in
W94 and cannot provide $m$ identifications.  The facts that the
multiplets do change and the 2006 combinations do not conform to
theoretical predictions leads us to expand our investigation to
include archival observations of GD358.  In the following, we will
examine the structure of GD358's multiplets in detail over timescales
of years.

\section{Multiplet Structure Change and Complexity}

Each multiplet in GD358's FT should represent a quantized g-mode
pulsation, described theoretically by a spherical harmonic of index
$l$ and overtone $k$, which has been split into multiple components
($2l+1$) by the star's rotation.  As component frequencies are
determined by the star's structure and rotation rate, we would expect
the multiplet structure to remain stable over long time periods. The
classic example is the DO pulsator PG1159-035, which exhibits
prototypical triplets, corresponding to $l=1$, and quintuplets,
corresponding to $l=2$ (see Figures~5 and 6 in Winget et al. 1991).

W94 based the first asteroseismological analysis of GD358 on the
identified modes and multiplet structure in the 1990 data set.  The
frequency splittings within the 1990 multiplets range from 3.5 to 6.5
\muHz, vary as a function of the radial number $k$, and are not always
symmetric with respect to the central frequency. W94 interpreted the
trend with $k$ as differential rotation within the star, and used the
splitting asymmetry to determine an average magnetic field strength of
$1300\pm300$ G. If we assume that GD358's structure is constant in
time, then we would expect to find similar multiplet structure in
other observing seasons, including XCOV25.

\begin{figure*}
 \includegraphics[trim= 0 80 70 60 ,clip,width=0.95\textwidth,angle=0]{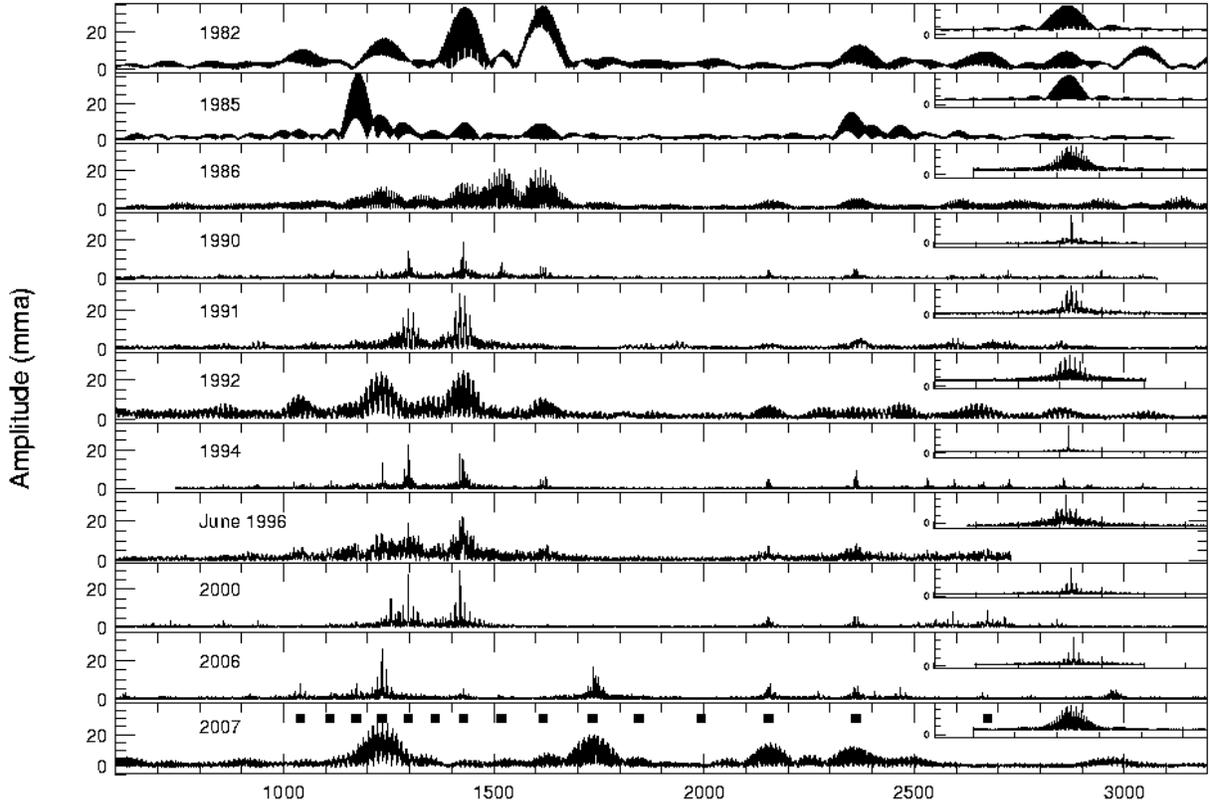}
 \caption{FTs of GD358 from 1982 to 2007, with each spectral window 
   given in the upper right corner of each panel. Each panel spans
   2800 \muHz. The filled squares in the bottom panel mark the
   predicted locations of GD358's pulsation modes ($l$=1).  GD358
   changes its distribution of amplitudes of the excited modes, but
   the modes are consistently in the same general locations.  High $k$
   modes exhibit the largest variation in amplitude. Note the
   appearance of $k=12$ in 2006. }
\end{figure*}

Figure~10 displays FTs for seasons spanning GD358's 1982 discovery to
2007. The 1990 (W94), 1991, 1994, 2000 (K03), and 2006 FTs are from
WET runs, while the other FTs are single site, obtained from McDonald
Observatory or Mt. Cuba Observatory (2007 season). The large amplitude
peaks are consistently found between 1000-1800 \muHz, and individual
modes are close to predicted values for $l=1$ (the filled boxes in the
last panel of Figure~10 marks the periods of the best fit model from
Metcalfe et al. 2003).  Using the 2006 data set, we calculated an
average period spacing of 38.77 s using the measured periods of the
lowest and highest $k$ modes.  This is consistent with results from
previous observations (Table~4), demonstrating that GD358's general
internal structure is largely unchanged.

\begin{figure}
 \includegraphics[width=1.00\columnwidth,angle=0]{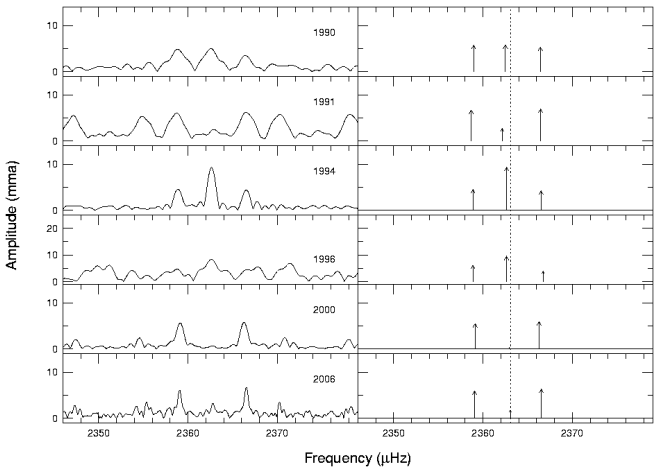}
 \caption{Multiplet structure associated with the $k=8$ mode in GD358
   from different observing seasons.  The left panels
   plot the original FTs, and the right panels give the
   prewhitening results. The largest 1$\sigma$ frequency error bar is 0.06 \muHz (1996).
   Each panel spans 18 \muHz.  The horizontal
   line through each right panel gives the noise level for that
   season, and the dotted line marks the 2006 detected frequency.}
\end{figure}

\begin{figure}
 \includegraphics[width=1.00\columnwidth,angle=0]{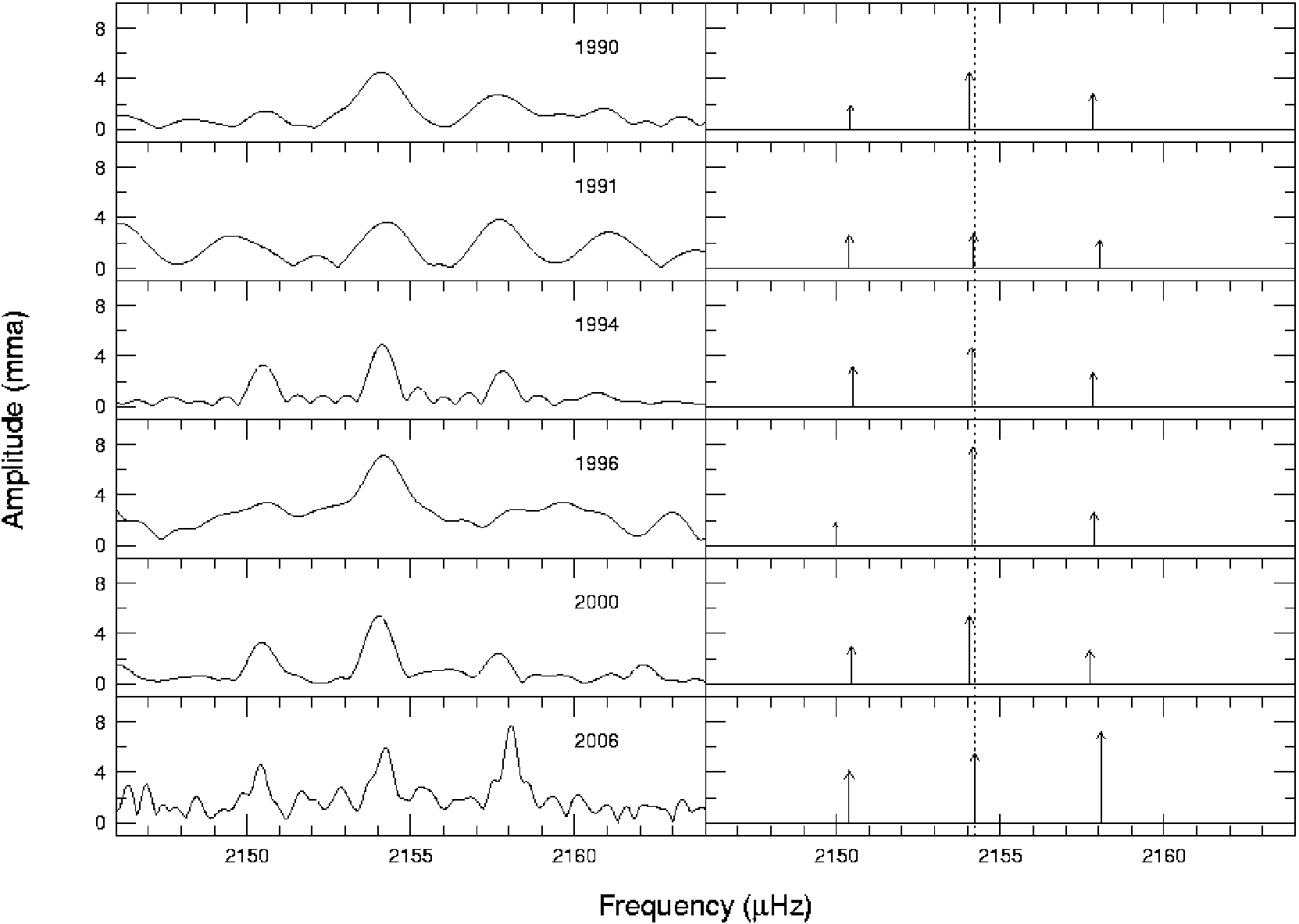}
 \caption{Multiplet structure associated with the $k=9$ mode in GD358
   from different observing seasons.  The left panels
   plot the original FTs, and the right panels give the
   prewhitening results. The largest 1$\sigma$ frequency error bar is 0.06 \muHz (1996).
   Each panel spans 18 \muHz.  The horizontal
   line through each right panel gives the noise level for that
   season, and the dotted line marks the 2006 detected frequency.}
\end{figure}

\begin{figure}
 \includegraphics[width=1.00\columnwidth,angle=0]{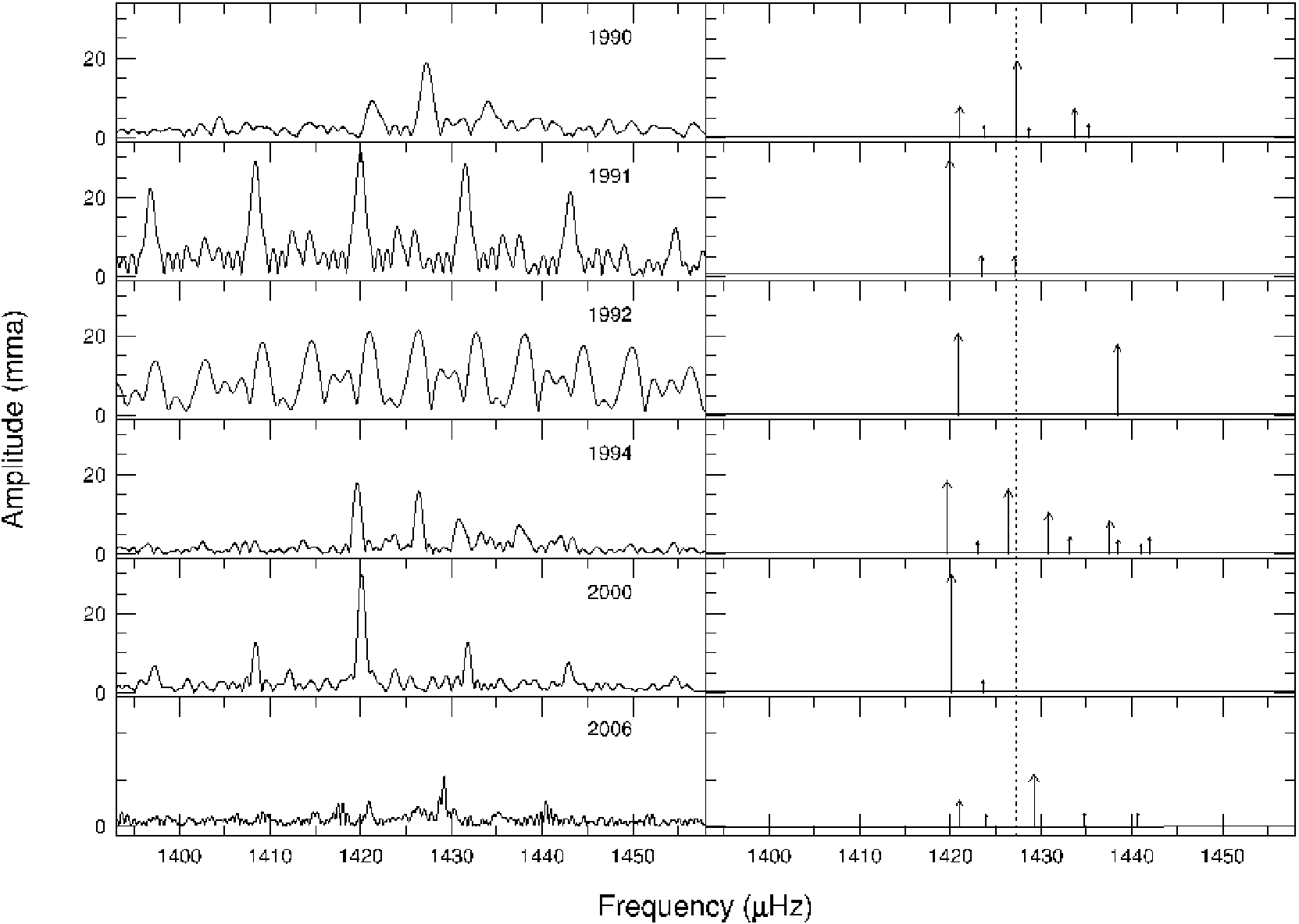}
 \caption{Multiplet structure associated with the $k=15$ mode in GD358
   from different observing seasons.  The left panels
   plot the original FTs, and the right panels give the
   prewhitening results. The largest 1$\sigma$ frequency error bar is 0.06 \muHz (1992).
   Each panel spans 65 \muHz.  The horizontal
   line through each right panel gives the noise level for that
   season, and the dotted line marks the 1990 detected frequency.}
\end{figure}

\begin{figure}
\includegraphics[width=1.00\columnwidth,angle=0]{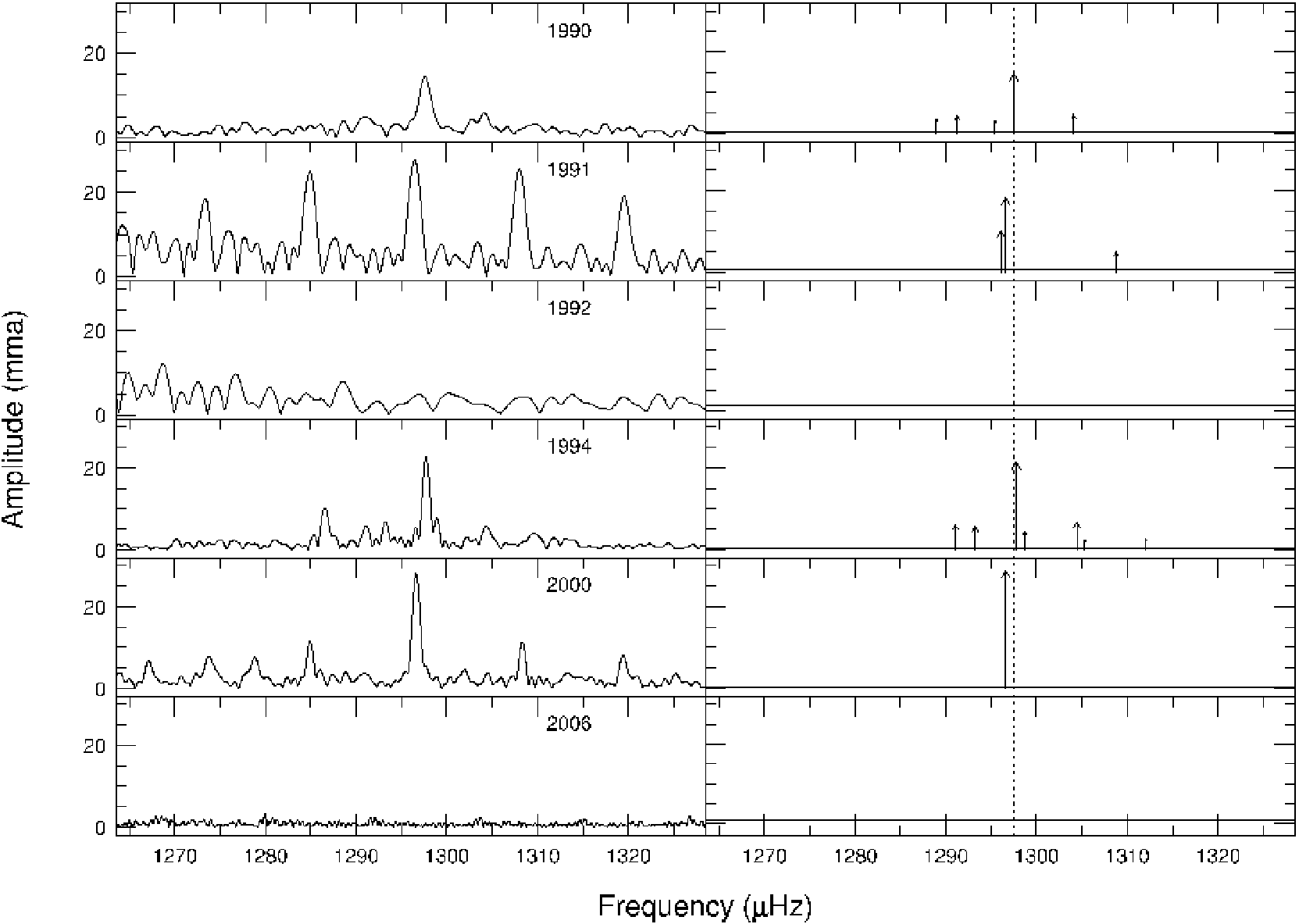}
 \caption{Multiplet structure associated with the $k=17$ mode in
   GD358 spanning 1990 to 2006.  The left panels present the yearly
   FTs, and the right panels represent the prewhitening results. The
   largest 1$\sigma$ frequency error bar is 0.06 \muHz (1992).  Each
   panel spans 65 \muHz.  The horizontal line through each right panel
   gives the noise level for that season, and the dotted line marks
   the 1990 detected frequency. The 2006 peak is difficult to detect
   due to the y-scale.}
\end{figure}

We extracted the $k=8$, $9$, $15$, and $17$ multiplets from the 1990,
1991, 1992, 1994, 1996, 2000, and 2006 seasons, using the techniques
discussed in Section 3. The multiplets are plotted in Figures~11, 12,
13, and 14.  The low-order $k=8$ and $9$ modes consistently display
the triplet structure expected for $l=1$ g-modes in the limit of slow
rotation (Figures~11 and 12).  While the component amplitudes vary from
season to season, the frequencies and splitting within the triplets
remain roughly the same.  The story changes for the $k=15$ and $17$
multiplets (Figures~13 and 14). The multiplet structure of these modes
exhibit dramatic changes from year to year.  If rotational splitting
is the dominant mechanism producing the high $k$ multiplets, we expect
them to share the same splitting with other detected modes within the
same data set. We have shown that this is not the case (Figure~7).

\begin{figure}
 \includegraphics[width=1.00\columnwidth,angle=0]{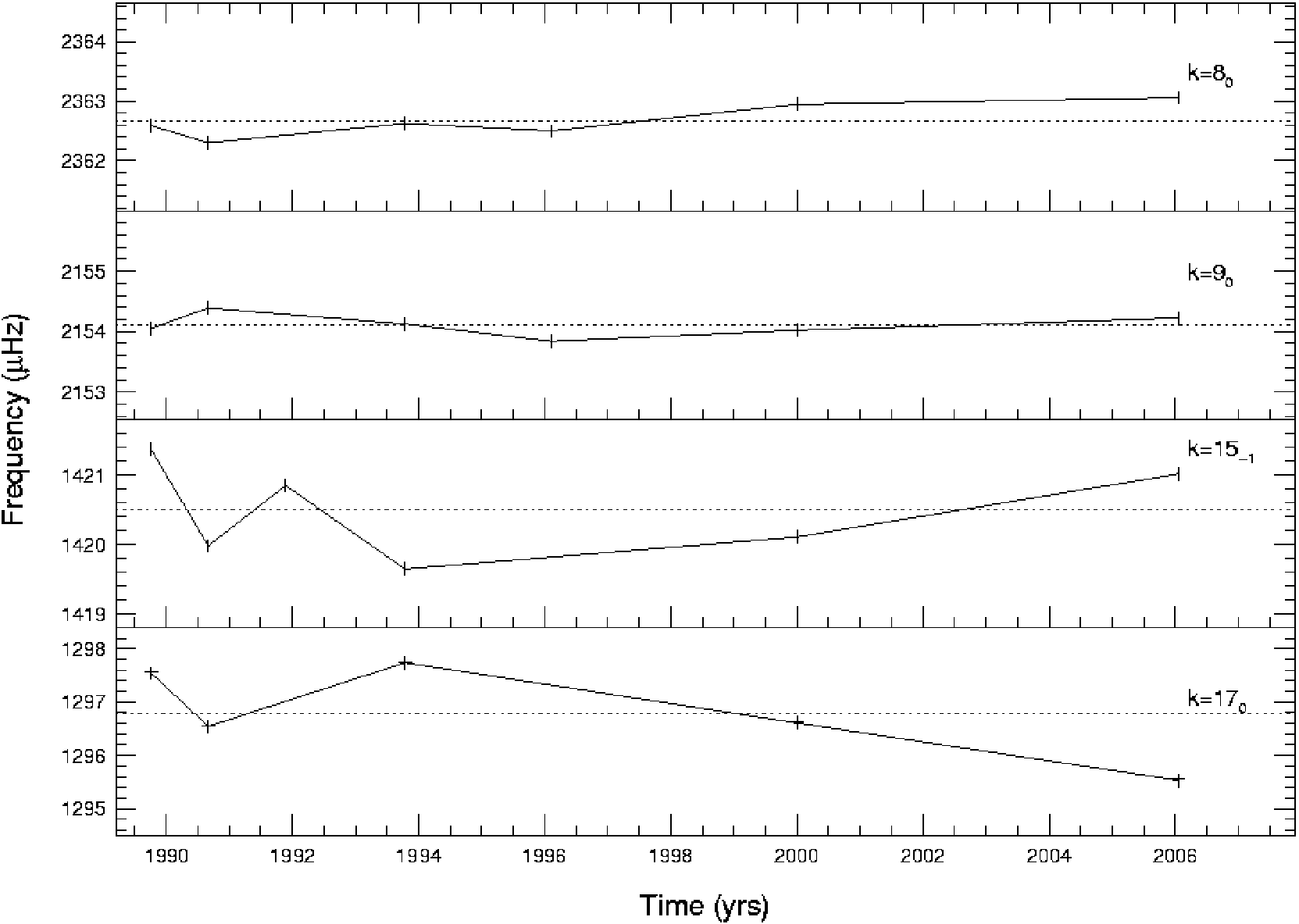}
 \caption{Frequency of the largest component for the $k$=17, 15, 9, 
   and 8 modes. For $k$=17, 9. and 8, the $m=0$ component was
   selected.  For $k=15$, the $m=-1$ mode was detected more often at
   larger amplitude.  While all modes exhibit some variation in
   frequency, the high $k$ modes wander by over 30 $\sigma$. The
   dotted lines mark the reported 1990 frequencies (W94).}
\end{figure}

We now turn our attention to the frequency stability of the $k$=8, 9,
15, and 17 mode components. Both W94 and K03 report small but
significant frequency changes within the identified multiplets, and
Figures~11--14 support this finding. For $k=8$, 9, and 17 we select the
component identified as $m=0$ in W94.  For $k=15$, however, we choose
$m=-1$ since a peak was present near this frequency more often than
the $m=0$ frequency. Figure~15 plots the frequencies of the identified
peaks versus time.  In all cases, the frequencies shift by several
times the statistical errors, where the average error for all the data
sets is $\sigma=0.06$ \muHz\.  The high $k$ frequencies wander by over
30 $\sigma$, much more than the detected 8 $\sigma$ changes for $k=8$
and 9. While these variations are small, they point to very definite
changes in the $g$-mode propagation cavity in this star.

Summarizing our results, we find a clear difference between high and
low $k$ modes.  The high $k$ multiplets are highly variable in
amplitude and frequency, and exhibit large changes in multiplet
structure.  The multiplets within these modes appear to be almost
randomly present with the $m=0$ component not necessarily preferred.
The low $k$ modes typically show a stable triplet structure, but the
amplitudes of the components are moderately variable.  All the
multiplets exhibit significant wanderings in frequency, with the
frequency changes observed in the high $k$ modes being much larger
than those seen in the low $k$ modes.

\section{Implications for GD358}

GD358's 2006 FT contains 27 independent frequencies distributed in 10
excited modes.  The locations of the modes agree with previous
observations and predicted values for $l$=1 $g$-mode pulsations.  The
average period spacing $\Delta P_{av}$, calculated using the lowest
$k$ and highest $k$ modes, is identical from 1990 to 2006 (Table~4).
As $\Delta P_{av}$ is fixed by stellar mass and temperature, GD358's
basic physical properties have remained unchanged over a timescale of
years, as expected.  The changes we observe in GD358's pulsation
spectra and multiplet structure are not due to gross changes in
structure or temperature.

We have shown that the standard model of a pulsating white dwarf in
the limit of slow rotation cannot adequately explain the observations
of GD358's multiplet structure and behavior. The fine structure of the
high $k$ modes is inconsistent with the expectations of slow
rotational splitting, the amplitudes of multiplet components for all
$k$s cannot be explained using simple viewing arguments, and the
observed frequency variations imply changes in the $g$-mode
propagation medium that are much more rapid than expected from
evolutionary changes.

The qualitative behavior of the low-$k$ and high-$k$ modes can be
explained in the context of a general model. $g$-modes are standing
waves of buoyancy in a spherical cavity, and they are thought of as
superpositions of traveling waves bouncing back and forth between an
inner and an outer turning point. These turning points depend on the
mode's period: short period (low $k$) modes have deeper outer turning
points than long period (high $k$) modes, so high $k$ modes sample the
outer regions of the star more than low $k$ modes.  If we introduce an
additional, non-spherical structure in the star's outer layers, due to
magnetic fields or convection, for example, the effect should be much
larger on the high $k$ modes.  In effect, GD358 is telling us where
this perturbation to its structure must be, but it is still incumbent
upon us to diagnose the actual cause.

We do find a clear difference in the photometric behavior of GD358's
high and low $k$ modes, leading us to consider a connection between
the outer layers and its pulsation changes. The following discussion
requires that we draw into the mix a remarkable sequence of events
that occurred over $\approx$30 days in 1996.  We follow Castanheira et
al. (2005) in calling the event the {\it sforzando}.  In classical
music, {\it sforzando} is an abrupt change in the character of music,
usually accompanied by an increase in volume.  We begin with a
description of the {\it sforzando}.  We will examine the implications
and connections between our asteroseismic results, the {\it sforzando}
and two possibilities that may affect GD358's photometric behavior: a
surface magnetic field and the convection zone.

\subsection{GD358 in 1996}

\begin{figure}
 \includegraphics[width=1.0\columnwidth,angle=0]{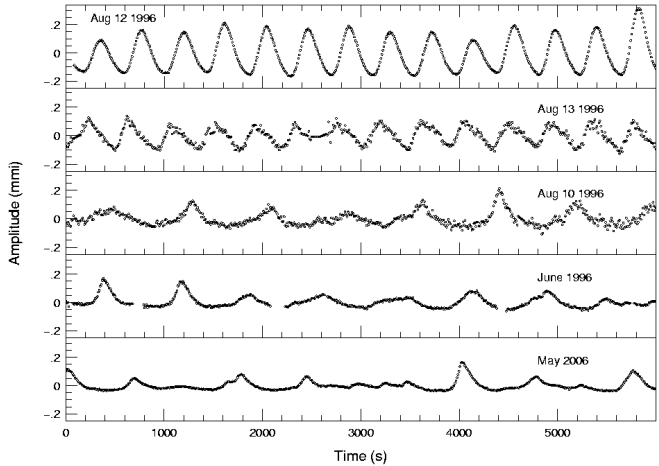}
 \caption{Light curve of GD358 on Aug 12 (an-0034, top panel) and 
   Aug 13 (suh-056, second panel), during the 1996 {\it sforzando}.
   On Aug 10 1996 (suh-0055, third panel), the light curve was
   typically nonlinear, as it was in June 1996 (fourth panel), and May
   2006 (bottom panel)}.
\end{figure}

In June 1996, GD358's light curve (third panel, Figure~16) was
typically nonlinear, with a dominant frequency of 1297 \muHz\ 
corresponding to $k$=17, and a peak to peak intensity variation of
$\approx$15\%. The $k$=9 and 8 modes were present with amplitudes of 8
and 9 mma, respectively. On August 10 (suh-0055, 2nd panel Figure~16),
the light curve was again typical, with a 20\% peak to peak intensity
variation, and main frequencies consistent with previously observed
high $k$ modes. The $k$=9 and 8 modes had upper amplitude limits of 8
mma.  On August 12, 27 hours later, GD358 exhibited completely altered
pulsation characteristics (an-0034, top panel Figure~16). The lightcurve
was remarkably sinusoidal, with an increased peak to peak intensity
variation of $\approx$50\%.  The $k$=8 mode dominated the lightcurve
with a amplitude of 180 mma, the highest ever observed for GD358. The
$k$=9 mode was also present, with an amplitude of $\approx$20 mma. The
high $k$ modes, dominant 27 hours prior, had upper detection limits of
7 mma. Over the next 24 hours, the $k$=8 mode began to decrease in
amplitude (Figure~17). Between Aug 13 and 14, $k$=9 grew in amplitude
as $k$=8 decreased, reaching a maximum of $\approx$57 mma, and then
began to decrease as the $k$=8 mode continued to dissipate. By August
16, a mere 96 hours after the {\it sforzando} began, GD358 was an
atypically low amplitude pulsator, with power detected in the $k$=9
and 8 modes at amplitudes of $\approx$7 mma, and upper limits of less
than 5 mma for the high $k$ modes. By September 10, less than a month
after the start of the {\it sforzando}, the high $k$ modes had
reappeared, with the dominant mode at $\approx$1085 \muHz\ at an
amplitude of 29 mma.

\begin{figure}
  \includegraphics[trim= 20 50 60 60,clip,width=1.0\columnwidth,angle=0]{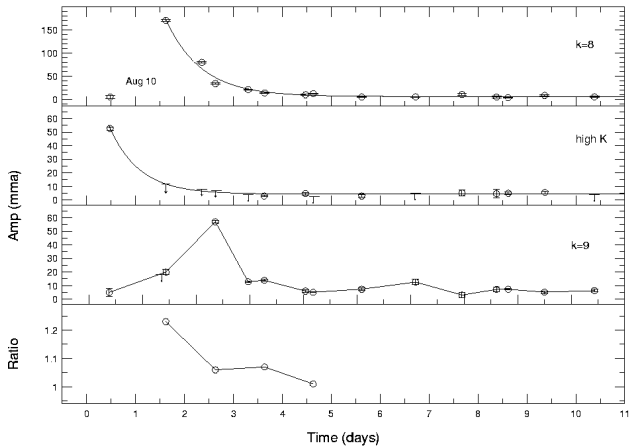}
  \caption{The first three panels give the measured amplitudes for
    $k$=8 (top panel), the high $k$ modes, and $k$=9 during the 1996
    {\it sforzando}. The solid lines give fits using a simple
    exponential decay model.  For $k$=8 and the high $k$ modes, the
    damping time was 1.4 days. Note that the ``high $k$'' points do
    not correspond to any particular $k$ value, but give the highest
    amplitude (or upper limit) between 1000 and 1500 \muHz at that
    time. The $k$=9 mode grew in amplitude as $k$=8 decayed. The last
    panel gives the ratio between GD358 and its comparison star from
    the McDonald 2.1m photometry. Note the change in y-axis units.  }
\end{figure}

The only observations we have during the maximum {\it sforzando} are
the PMT photometry (Table~5) observations. Differential photometry
does not measure standard stars.  However, we can calculate the ratio
between the average counts from GD358 and its comparison star,
assuming the same comparison star was used each night, and accounting
for factors such as extinction and sky variation.  In the following,
we calculate ratios based on sky subtracted counts 
obtained within $\pm1.5$ hrs of the zenith.  We focus on the McDonald
2.1m runs an-0034, an-0036, an-0038, and an-0040.  The ratio between
GD358 and the comparison star decreased from 1.23 during the height of
the {\it sforzando} (an-0034, Aug. 12) to 1.01 on Aug. 15 (an-0040).
We also examined observations from Mt. Suhora (Table~5). Suh-0055 was
obtained $\approx$24 hrs before the {\it sforzando}, and the ratio
between GD358 and the comparison star was 0.11.  Suh-0056 took place
18 hours after the {\it sforzando} maximum but before the pulsations
had completely decayed.  For suh-0056, the ratio between GD358 and the
comparison star increased to 0.16. For the remaining Mt. Suhora
observations, obtained during the following days, the ratio between
GD358 and the comparison star decreased to 0.12.  Observation logbooks
at Mt. Suhora demonstrate that the same comparison star, but not the
same one used at McDonald, was used each night (Zola 2008).  While we
can't pin down any exact numbers due to the nature of differential
photometry, the observations of GD358 are consistent with a flux
increase in the observed bandpass during the maximum {\it sforzando}.

\subsection{Convection and GD358}

The onset of convection goes hand in hand with the initiation of
pulsation in both hydrogen and helium white dwarfs (Brickhill 1991, Wu
2000, Montgomery 2005). A primary goal of XCOV25 is to use GD358's
nonlinear lightcurve to characterize its convection zone, improving
the empirical underpinnings of convective heat transport theory.  Our
asteroseismological investigation points to evidence that something in
the star's outer surface layers is influencing its pulsation modes.
The convection zone lies in the outermost regions of GD358, so we are
led to consider the relationship between the star's convection zone
and its photometric behavior, both typical and during the {\it
  sforzando}.

In general, theoretical studies of convective energy transport are
based on the mixing length theory (MLT) (Bohm-Vitense 1958).
Originally intended to depict turbulent flows in engineering
situations, MLT enjoys success in describing stellar convection (Li \&
Yang 2007).  It remains, however, an incomplete model with unresolved
problems. The actual mixing length is not provided by the theory
itself, but is defined as ($\alpha{H_p}$), where $H_p$ is the pressure
scale height, and $\alpha$ is an adjustable variable.  This adjustable
parameter undermines the 
ability of theoretical models to make useful predictions.
Advances in convection theory include stellar turbulent
convection (Canuto \& Mazzitelli 1991, Canuto et al. 1996), which
establishes a full range of turbulent eddy sizes.  Work is currently
underway employing helioseismology to compare MLT and turbulent
convection models (Li \& Yang 2007).  Our work with convective light
curve fitting will also provide important tests for convection theory.

Physical conditions in white dwarf atmospheres differ by orders of
magnitude from those in envelopes of main sequence stars like the Sun.
For GD358, models indicate that the convection zone is narrow, with a
turnover time of $\approx$1 second and a thermal relaxation timescale
of $\approx$300 seconds (Montgomery 2008).  The pressure scale height
is small, limiting the vertical height of the convective elements.
Montgomery \& Kupka (2004), employing an extension of stellar
turbulent convection, calculate that 40-60\% of GD358's flux is
carried by convection, depending on the adopted effective 
\linebreak 
 \begin{deluxetable*}{ccccccc}
 \tablecolumns{3}
 \tablewidth{0pc}
 \tablecaption{GD358 August 1996}
 \tablehead{
 \colhead{Run} &\colhead {CH1} &\colhead{CH2} &\colhead{CH1/CH2} & Time & Date & Length\\
 \colhead{} &\colhead{GD358} &\colhead{Comp} & & (BJED) &  & (hrs)
 }
 \startdata
 McDonald \\
 an-0034 & 343727 & 279268 & 1.23 & 2450307.6170160 & Aug 12 & 4.7\\
 an-0036 & 308308 & 290898 & 1.06 & 2450308.6309639 & Aug 13& 3.2\\
 an-0038 & 269937 & 251923 & 1.07 & 2450309.6362315 & Aug 14 & 3.5 \\
 an-0040 & 250813 & 248479 & 1.01 & 2450310.6294625 & Aug 15& 3.5 \\
 Mt. Suhora \\
 suh-0055 & 12786 & 114746 & 0.11 & 2450306.4793479& Aug 10& 4.8\\
 suh-0056 & 29408 & 181597 & 0.16 & 2450308.3533174& Aug 12& 3.0\\
 suh-0057 & 20271 & 158035 & 0.12 & 2450309.3015281 & Aug 13& 5.9\\
 suh-0058 & 15974 & 130814 & 0.12 & 2450310.4729899& Aug 14& 1.0\\
 suh-0059 & 15434 & 127672 & 0.12 & 2450314.3776294 & Aug 18& 3.0\\
 \enddata
 \tablecaption{Observations made with McDonald 2.1m.  The same 
 comparison star was used in each instance.}
 \end{deluxetable*}

\hspace*{-2.2em}
temperature.

Convective lightcurve fitting is based on the assumption that the
convection zone, by varying its depth in response to the pulsations,
is responsible for the nonlinearities typically
observed in GD358's lightcurves (Montgomery 2005). During the
\emph{sforzando}, within an interval of 27 hours, GD358's lightcurve
became sinusoidal, arguing that the convection zone was inhibited, at
least as sampled by the k=8 mode.  For example, if this large
amplitude mode was the m=0 member of the triplet, then its brightness
variations would be predominantly at the poles and not the equator.
Thus, all that is required to produce a sinusoidal lightcurve is a
mechanism to inhibit convection near the poles.

How could convection in any star be decreased on such short
timescales?  The obvious method, but not necessarily the most
physical, is to raise GD358's effective temperature, forcing it to the
blue edge of the instability strip, and decreasing convection.  Such a
temperature increase has the observable consequence of increasing the
stellar flux.  A flux increase of 20\% in our effective photometric
bandpass would produce an equivalent change in the bolometric flux of
$\approx$40\%, corresponding to a temperature \ change of
$\approx$2200 K.  The {\it sforzando} photometry is consistent with
such an increase in flux, and Wiedner \& Koester (2003) find that
$\rm{T_{eff}}\approx27,000$K is required to simulate GD358's light
curve during the maximum {\it sforzando}. Castanheira et al. (2005)
use FOS spectroscopy of GD358 obtained on 1996 August 16 to determine
$\rm{T_{eff}}=23900\pm1100$K, a number not inconsistent with previous
observations (Beauchamp et al. 1999).  However, the FOS spectra were
acquired after the pulsations had greatly decreased in amplitude
(Figures~16 \& 17). It is possible that GD358 had cooled by this point.
Recent theoretical evidence also indicates that simply ``turning off''
convection would have minimal observable effects on GD358's spectrum
(Koester 2008).
 
The connection between convection and pulsation during the {\it
  sforzando} may also depend on the origin of the event.  Was the {\it
  sforzando} a short term change, such as a collision, or a more
global event intrinsic to GD358?  The growth and dissipation of modes
during the {\it sforzando} could be connected to pulsation growth
timescales, or may be governed by a completely different mechanism.
Pulsation theory predicts that growth rates for high $k$ modes are
much larger than for low $k$ modes.  This simply means that high $k$
modes are easier to excite and have growth times (and decay times) of
days rather than years (Goldreich \& Wu 1999).  This makes sense for
GD358, where we typically find the high $k$ modes to be much more
unstable in amplitude over time than $k$=9 and 8 (Figure~10).  Figure~17
plots the amplitudes for the $k$=8, $k$=9, and ``high $k$'' modes
during the {\it sforzando}. We note that the ``high $k$'' points do
not correspond to any particular $k$ value, but give the highest
amplitude (or upper limit) between 1000 and 1500 \muHz.  We fit the
observed amplitudes for $k$=8 and the high $k$s with a simple
exponential decay model $A(t)=A_{0}e^{-t/\tau_d}+c$, where A(t) is the
amplitude at time t and $\tau_d$ is the damping constant.  Our fits
give a damping timescale of $\tau_{d}=1.5\pm0.2$ days for both the
$k$=8 mode and the high $k$s. The $k$=9 mode, which actually grew in
amplitude later, as $k$=8 decayed, has an upper limit of
$\tau_{9^d}<1.1$ days.  We only have a few points measuring the growth
timescale ($\tau_{g}$) of $k$=8 and 9, so we place an upper limits for
$k$=8 of $\tau_{8^g}\le1.2$ days.  For $k$=9, we find an upper limit
of $\tau_{9^g}\le2.2$ days.  These timescales are all roughly
equivalent, are all much longer than the dynamical timescale, and not
distributed as predicted by theoretical pulsation growth rates. The
results for $k$=9 and 8 are much shorter than the expected pulsation
damping timescale of about 2 months (Montgomery 2008a), arguing that
the source of the {\it sforzando} was capable of interacting with the
pulsations to produce changes on faster timescales than theoretical
growth rates.

Our purpose in entering this discussion of the {\it sforzando} was to
provide insight into connections between convection and pulsation in
GD358. We find evidence that convection diminished during the {\it
  sforzando}. The decreased convection was accompanied by the rapid
disappearance of GD358's high $k$ modes and the transfer of energy to
the $k$=8 and 9 modes.  The photometry is consistent with a
temperature increase, which would inhibit convection.  The sudden
transfer of energy from the high $k$ modes to $k$=9 and 8 argues that
a mechanical (radial or azimuthal) or thermal structural change
altered GD358's outer layers, modifying mode selection. We find that
the growth/damping timescales during the {\it sforzando} for the $k$=9
and 8 modes do not agree with the expectations of pulsation theory,
and may be governed by a completely different mechanism.

Magnetic fields have long been known to be capable of inhibiting
convection.  For example, localized fields associated with sunspots in
the Sun are observed to inhibit convection in the surrounding
photosphere (Biermann 1941, Houdek et al. 2001).  Winget (2008)
estimates that the magnetic decay time at the base of the convection
zone is $\approx$3 days. In the next section, we investigate the
possible implications of magnetic fields on GD358 and its pulsations.

\subsection{Magnetic Fields}

Our current understanding of magnetic fields in stars like the Sun
invokes the dynamo mechanism, which is believed to function in the
narrow region where rotation changes from the latitudinal rotation of
the outer convective layers to spherical rotation in the radiative
zone (Morin et al. 2008).  The solar magnetic field consists of small,
rapidly evolving magnetic elements displaying large scale
organization, with a cycle time of 22 years during which the polarity
of the field switches. Localized fields, with strengths hundreds of
times stronger than the global field, exert non-negligible forces on
charged particles, influencing convective motion. Solar pulsations are
observed to vary in frequency with the solar cycle in these regions
(Schunker \& Cally 2006). Although the exact mechanism is uncertain,
the frequency shifts are interpreted as representing change in the
Sun's internal structure or driving (Woodward \& Noyes 1985, Kuhn
1998, Houdek et al. 2001).

A pulsating white dwarf represents an extreme environment.  The
atmosphere and the convection zone are thin, the surface gravity
orders of magnitude higher, and differential rotation may or may not
play a role (Kawaler et al. 1999). The magnetic field in a typical
white dwarf is confined to the nondegenerate outer layers.  The field
may dominate near the surface, but deeper in the atmosphere, gas
pressure will prevail.  Drawing an analogy with the Sun, a surface
magnetic field in GD358 probably has an associated cycle time similar
to the solar cycle. Hansen et al. (1977) and Jones et al. (1989)
discuss this topic from a theoretical perspective, predicting cycle
times ranging from 2 to 6 years.

What are the observable consequences of a magnetic field on white
dwarf pulsations?  Jones et al. (1989) finds that, in the presence of
a weak field, defined as strong enough to perturb but not dominate
mass motions, multiplet frequencies are increased with respect to the
central mode in a manner proportional to $m^2$. The $m$=0 mode is also
shifted, unlike the case for rotation. Each $g$-mode pulsation samples
the magnetic field at a different depth.  A logical expectation is
that, for low $k$ modes with reflection points below the convection
zone, we can treat the magnetic field as a perturbation, while for
high $k$ modes such a treatment is not valid.

\begin{figure}
 \includegraphics[width=1.00\columnwidth,angle=0]{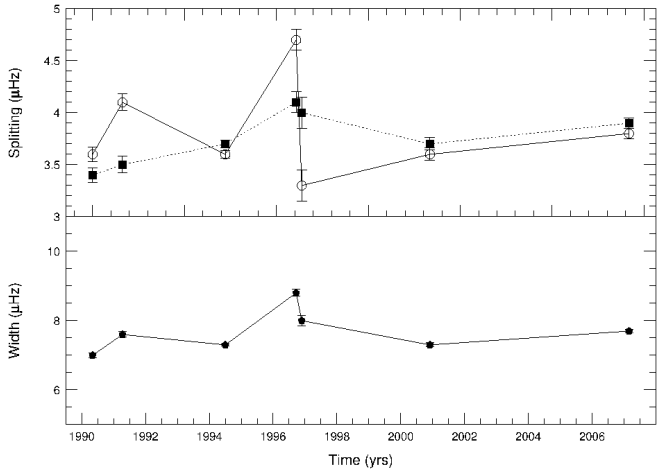}
 \caption{Multiplet structure within the $k$=9 multiplet.  The x-axis 
   is time in years, and the y-axis is in \muHz. The bottom panel
   plots the width of the entire multiplet over time.  The top panel
   examines the $m=0,+1$ (dotted line) and $m=0,-1$ (solid line).  }
\end{figure}

\begin{figure}
  \includegraphics[trim= 50 60 65 70,clip,width=1.00\columnwidth,angle=0]{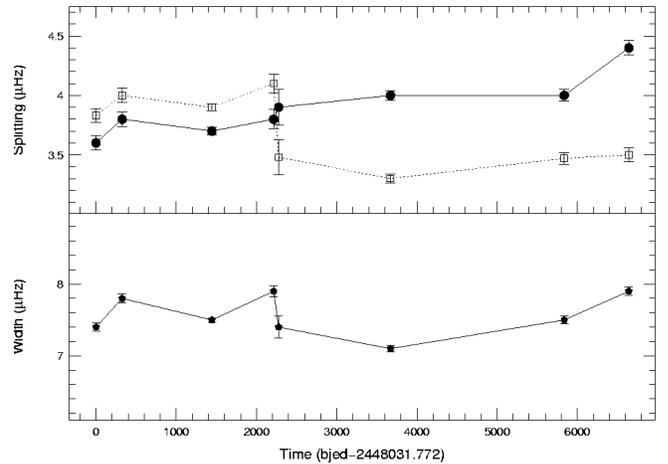}
\caption{Multiplet structure with $k$=8 multiplet. The x-axis is 
  time in years, and the y-axis is in \muHz. The bottom panel plots
  the width of the entire multiplet over time.  The top panel examines
  the $m=0,+1$ (dotted line) and $m=0,-1$ (solid line).  }
\end{figure}

Our best candidates to investigate the influence of a magnetic field
on multiplet structure are GD358's $k$=9 and 8 modes.  These two modes
are consistent with rotationally split triplets and we can examine
their multiplet structure in detail over 16 years of observations
(Figures~18 and 19).  From 1990 to 2006, $k=9$ had an average
multiplet width of $7.7\pm0.1$ \muHz, while $k=8$ is similar, at
$7.5\pm0.1$ \muHz.  Figures~18 and 19 show that the complete multiplet
width and the splittings with respect to the central mode wander up to
0.5 \muHz\ from the average values.  Both Figures~18 and 19 reveal a
dramatic change during the {\it sforzando}.  In June 1996, both
multiplets increased in total width by $\approx$1\muHz. Prior to
August 1996, the retrograde splitting (0,-1) for $k$=8 was
consistently smaller by $\approx$0.2 \muHz than the prograde splitting
(0,+1). After August 1996, the retrograde splitting became
consistently larger by $\approx$0.5 \muHz\ than the prograde splitting
and has remained so through 2007, at the resolution of our
observations. The case is not so straightforward for $k$=9, but this
mode clearly shows a large change in 1996, especially for the (0,-1)
splitting (Figure~18).  In the previous section, we questioned whether
the {\it sforzando} was a short term event or a global, enduring
change.  This analysis of $k$=9 and 8 points toward a long term change
in GD358's resonance cavity.  The changes we observe in the multiplet
structure could be explained by a change in magnetic fields.

A magnetic field also introduces an additional symmetry axis. Our
analysis of GD358's harmonics and combination frequencies hints that
something is disrupting its azimuthal symmetry. It is conceivable that
high $k$ modes would align more or less with the magnetic field, while
the low $k$ modes align with the rotation axis.  Wu (2001) and Yeates
et al. (2005) do not explore the possibility of multiple axes or the
effects of a magnetic field on nonlinear mixing by the convection
zone.  Future work could explain why GD358's modes do not combine as
expected by simple theory.

\begin{figure}
 \includegraphics[width=1.00\columnwidth,angle=0]{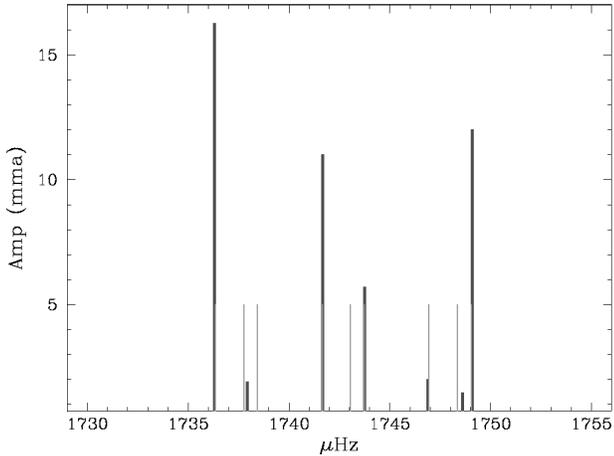}
 \caption{Fit to $k=12$ components assuming the dominant splitting 
   mechanism is a magnetic field using the model of Unno et al. (1979)
   rather than rotation.  The black lines are the observed multiplet
   components, including several below our detection limits that are
   not listed in Table~2.  The light gray lines are the theoretical
   predictions. Theoretical predictions overlap the 4 largest
   amplitude peaks, although they are difficult to distinguish in the
   figure. }
\end{figure}

The presence of the $k$=12 mode offers a final hint of a possible
surface magnetic field. The second largest mode in the 2006 FT, $k$=12
appears in a region that is normally devoid of significant power. If,
for this mode, the magnetic field can no longer be treated as a
perturbation, we would expect a multiplet with $(2l+1)^2$ components,
as opposed to the 3 expected from rotation (Unno et al. 1979).  For
$l$=1, this corresponds to 9 magnetic components.  We do find 6 peaks
above our $4\sigma$ limit, with 3 additional peaks slightly just our
criteria.  Using the magnetic model of Unno et al. (1797), we are able
to match the frequencies of the $k$=12 mode components, but not the
amplitudes (Figure~20).

\section{Summary and Conclusions}

GD358 is the best studied of the DB pulsators, yet our work shows that
this object is by no means completely understood. The 2006 XCOV25
observations were obtained with the goal of using GD358's nonlinear
lightcurve to characterize its convection zone, but our initial
asteroseismological analysis of the data set reveals a great deal of
interesting information about GD358's pulsational behavior.  Our
investigation began with an analysis of GD358's 2006 XCOV25 FT.  We
explored the identified modes, combination frequencies, and multiplet
structure.  Difficulties in identifying the $m$ components of the 2006
mode multiplets, both directly and indirectly using the combination
frequencies, lead us to examine the multiplet structure in detail over
time. Our investigation expanded to include observations of GD358 over
the 24 years since its discovery, focusing on the multiplet structure,
frequency stability, and photometric behavior of GD358. We summarize
our results concerning GD358's pulsation properties:

\begin{itemize}
 
\item The 2006 FT contains 27 independent frequencies distributed in
  10 modes ($k$=21, 19, 18, 17, 15, 14, 12, 11, 9, and 8).  The
  dominant frequency is at 1234 \muHz\ ($k=18$) with an amplitude of
  24 mma.
\item We find significant power at $k$=12, a region of the FT
  previously devoid of significant power.
\item The frequency location of each mode in the 2006 FT is consistent
  with previous observations and theoretical predictions, assuming
  $l$=1.
\item The $k$=9 and 8 modes exhibit triplets in agreement with
  theoretical predictions for $l$=1 in the limit of slow rotation. The
  amplitudes and frequencies of the components exhibit some
  variability over time, but much less than the high $k$ modes.
\item Our analysis of GD358's high $k$ multiplets over time reveal
  that they are variable in multiplet structure, amplitude, and
  frequency.  The variability and complexity of the high $k$
  multiplets cannot be interpreted simply as $l$=1 modes in the limit
  of slow rotation.
\item We cannot provide $m$ identifications for most of the multiplet
  components in the 2006 FT, with the exception of $k$=9 and 8.
\item The 2006 FT contains a rich assortment of combination
  frequencies.  They are potential tools for identifying $m$ values
  and orientation for each mode.  However, their amplitudes do not
  agree with theoretical predictions from Wu (2001).
\item The $k$=9 and 8 multiplet amplitudes and presence/absence of
  harmonics cannot be explained by simple geometric viewing arguments,
  and argue that something is interrupting GD358's azimuthal symmetry.
\item The linear, sinusoidal shape of the lightcurve indicates that
  GD358's convection zone was diminished during the 1996 {\it
    sforzando}.
\item Photometry taken during the maximum {\it sforzando} is
  consistent with a flux increase in the effective bandpass.
\item We find damping/growth timescales during the {\it sforzando}
  that are not consistent with the expectations of pulsation theory.
\item Changes in the splittings of the $k$=9 and 8 multiplets indicate
  that some mechanism, perhaps a magnetic field, induced a long-term
  change in the multiplet structure of these modes during the {\it
    sforzando}.

\end{itemize}

Our investigation raises a number of interesting implications for our
understanding of the physics of GD358, many beyond the scope of this
paper, and some blatantly skirting the realm of speculation.  While we
cannot pretend that our investigation has unearthed unshakable
evidence thereof, we do find tantalizing indications pointing to
connections between convection, magnetic fields, and pulsation in
GD358.  We suggest future investigations:

\begin{itemize}
\item Theoretical investigation of the relationship/connection between
  magnetic fields, convection, and pulsation will increase our
  understanding of GD358's photometric behavior. The typical static
  model of a pulsating white dwarf is too limiting.  The dynamic model
  must accommodate the observed changes in pulsation frequencies and
  multiplet structure.
\item GD358's multiplets do not conform to theoretical expectations
  based on rotation.  Other mechanisms must be considered.
  Examination of the influence of a nonspherical asymmetry in the
  outer layers on the multiplet structure is required to understand
  GD358's multiplets. Identification of spherical degree by multiplet
  structure alone should be highly suspect in any large amplitude
  pulsator.
\item Further work on the theoretical aspects of combination
  frequencies should explore the effects of multiple symmetry axes and
  the effects of a magnetic field on nonlinear mixing by the
  convection zone.
\item Detailed investigation of the FOS spectrum obtained during the
  {\it sforzando} for possible metals may provide insight into the
  mechanism producing the event.
\item Theoretical predictions for growth and damping rates have been
  calculated using static models.  The timing of the growth and decay
  of $k$=8 and 9 during the {\it sforzando} indicates interaction
  between these two modes. A theoretical examination of growth rates
  in the presence of other modes, including the interaction of a mode
  with itself, is necessary to better understand both the growth and
  dissipation of modes during the {\it sforzando} and typical mode
  selection in all large amplitude DB and DA pulsators.
\item A detailed examination of behavior of white dwarf pulsators,
  both low and high amplitude, spanning both the DA and DB instability
  strips (and hence a range of convective depths) will improve our
  understanding of convection's role in mode selection.
\item Theoretical work is needed to improve the treatment of
  convection. Lightcurve fitting of GD358 and other pulsators is an
  important step towards that goal.
\item Continued observation of GD358 will better define its dynamic
  behavior.  We would like to observe another {\it sforzando}!
\end{itemize}

The practice and theoretical development of asteroseismology of GD358
and other pulsating stars continues to reward us with a rich
scientific return.  Our focus here on GD358 shows us that stellar
seismology can challenge our current paradigm of the interior and
behavior of pulsating white dwarfs.

\acknowledgments

The Delaware Asteroseismic Research Association is grateful for the
support of the Crystal Trust Foundation and Mt. Cuba Observatory. DARC
also acknowledges the support of the University of Delaware, through
their participation in the SMARTS consortium.  This work is further
supported by the Austrian Fonds zur F\"orderung der wissenschaftlichen
Forschung under grant P18339-N08.  We would like to thank the various
Telescope Allocation Committees for the awards of telescope time. We
also acknowledge the assistance of J. Berghuis, M. David, H. Wade, and
U. Burns with the Hawaii 0.6m.

{\it Facilities:} \facility{MCAO: 0.6m }, \facility{McD: 2.1m },
\facility{KPNO: 2.1m}, \facility{UH: 0.6m}, \facility{BOAO: 1.8m}, 
\facility{Lulin: 1.8m}, \facility{Beijing: 2.16m},
\facility{Maidanek: 1.0m}, \facility{Peak Terskol: 2.0m},
\facility{Moletai: 1.65m}, \facility{Mt. Suhora: 0.6m},
\facility{Konkoly: 1.0m}, \facility{Vienna: 0.8m},
\facility{T\"ubingen: 0.8m}, \facility{OHP: 1.93m}, \facility{NOT}, 
\facility{ING: Herschel}, \facility{LNA: 1.6m},
\facility{CTIO: 0.9m}, \facility{SOAR}

\end{document}